\documentclass[aps,11pt,footnote]{revtex4-2}
\usepackage{graphicx}
\usepackage{xcolor}
\usepackage{subfig}
\usepackage{amssymb}
\usepackage{lipsum}
\usepackage{physics}
\usepackage{mathrsfs}
\usepackage{MnSymbol}

\newcommand{\be}{\begin{equation}}
\newcommand{\ee}{\end{equation}}

\newcommand{\PR}[1]{\ensuremath{\left[#1\right]}}
\newcommand{\PC}[1]{\ensuremath{\left(#1\right)}}
\newcommand{\chav}[1]{\ensuremath{\left\{#1\right\}}}

\begin{document}
\title{Localized structures in three-field models: geometrically constrained configurations and the first-order framework{\vspace{0.5cm}}}

\author{D. Bazeia\footnote{ dbazeia@gmail.com}}
\author{G. S. Santiago\footnote{gss.santiago99@gmail.com}}

\affiliation{Departamento de Física, Universidade Federal da Paraíba, João Pessoa, Paraíba, Brazil}

\begin{abstract}

\vspace{0.8cm}

{\begin{center}{ABSTRACT}
\end{center}}

This work deals with models described by three real scalar fields in one spatial dimension. We study the case where two of the three fields engender kinematical  modifications, which respond as geometrical constrictions, that can be used to change the center and/or the tails of the kinklike configurations. An important advantage of our procedure is the construction of a method for the obtention of first-order differential equations that solve the equations of motion and give rise to stable localized structures. We illustrate the general procedure investigating several distinct examples, and suggesting some possibilities of applications of practical use, in particular, to the case of domain walls and skyrmions in magnetic material, collisions of kinks and to deal with braneworld scenarios having a warped five-dimensional anti de Sitter geometry, with a single extra dimension of infinite extent. 
\end{abstract}

\maketitle

\vspace{1cm}

\section{Introduction}

Nonlinearity plays an important role in physics in general. In high energy physics, in particular, nonlinearity is the source of many interesting phenomena. For instance, relativistic real and complex scalar fields, spinors and Abelian and non Abelian gauge fields may self interact and interact with one another, following standard and modified rules of coupling, to generate distinct scenarios of physical interest. They may contribute to generate localized structures that attain topological features that protect them against instability. 

 Among the several distinct types of localized structures, one may find kinks, vortices and magnetic monopoles, all having importance in distinct areas of high energy physics; see, e.g., Refs.  \cite{B1,B2,B3,B4} and references therein.  Here we shall restrict the study to the case of kinks \cite{baz}, which are perhaps the simplest of the topological structures, but we deal with models described by three real scalar fields, concentrating on the description of new features, that appear under the modification engendered by the systems. Before investigating three-field models, let us first remind that kinks have appeared in physics in a diversity of contexts, as in Refs. \cite{K1,K2,K21,K211,K3,K31,K4}, for instance, where models supporting kinklike configurations were deformed to produce new models together with their corresponding localized configurations \cite{K1}, to describe
 kinklike waves in a massive nonlinear sigma model with an $S^2$ sphere as its target manifold \cite{K2}. Kinks can also appear as Schrödinger kinks, in a model of a quantum phase transition, in which a topological defect can be in a non-local superposition, so that the order parameter of the system is in a quantum superposition of conflicting choices of the broken symmetry \cite{K21}, and also, in the structural phase transition from linear to the zigzag conformation in ion Coulomb crystals \cite{K211}, and
 to interpolate string solutions in a $SU(2)$ model in the presence of two Higgs fields \cite{K3}. They
  have also been experimentally realized as high-frequency topological modes (kink magnetoplasmons) in magnetoplasmonic devices featuring oppositely biased magnetic domains \cite{K31}, and as
thermally-activated  production of topological solutions, generating localized regions of bending in compressed filament bundles \cite{K4}.

There are several other possibilities, but here we shall focus on models described by three real scalar fields. Such models have been investigated before under distinct motivations, for instance, to study the entrapment of a network of domain walls \cite{31}, to describe three-field models 
 that lead to nontrivial cosmological behavior 
 \cite{32}, to manipulate the internal structure of Bloch walls \cite{33}, 
 to study inflation 
  with three fields with a specific hierarchical mass spectrum, when two fields act as inflatons and the third one is the spectator \cite{36},  
 to control the profile of thick brane in the five-dimensional scenario with an extra spatial dimension of infinite extent \cite{34},   
 and to enhance the power spectra from three-field inflation scenario \cite{35}.
 
 The main motivation of the present study is based on the recently uncovered phenomenon, which appears in a system of two real scalar fields, $\phi$ and $\chi$, in which the presence of the second field $\chi$ may contribute to modify the kinematics of the first field $\phi$. This was first discussed in \cite{Const}, and them further considered in \cite{sala,tail}. The presence of the second field to modify the kinematics of the first field, contributes very importantly to change the profile of the localized structure and this can be directly related to the experimental phenomenon uncovered sometime ago in magnetic materials \cite{bish}. As one can see, in \cite{bish} the authors reported that, using samples of magnetic materials constructed under the presence of a geometric  constriction, the magnetization may behave very differently from the standard case, inducing an internal modification in the localized configuration, having a two-kink profile theoretically described before in \cite{PRL}, and further studied in \cite{Const} under the presence of the kinematical modification. 

The organization of the present investigation is as follows. In the next Sec. \ref{II} we deal with the methodology, defining the general model and describing the main steps to get to the presence of first-order differential equations that solve the equations of motion and minimize the energy of the corresponding solutions. The first-order equations provide stable solutions and in Secs. \ref{III} and \ref{IV} we illustrate the general results with several examples, considering two distinct families of models. We close the work in Sec. \ref{V}, commenting on the main results and adding some perspectives of future investigations. 

\section{Generalities}\label{II}

In this work, we deal with models described by three real scalar fields in 1+1 spacetime dimensions. The main motivation is to study how the modification in the kinematics of some fields may induce new physical behavior in the fields themselves. This modification can be included in several different ways, but here it is due to the introduction of the functions $g(\psi)$ and $f(\chi,\psi)$ on the kinematic term of $\chi$ and $\phi$, respectively. In order to implement our investigation, we consider the Lagrange density in the form
\begin{equation}
    \mathcal{L}(\phi,\chi,\psi) =\frac{1}{2} f(\chi,\psi)\partial_{\mu}\phi\partial^{\mu}\phi +\frac{1}{2} g(\psi)\partial_{\mu}\chi\partial^{\mu}\chi + \frac{1}{2}\partial_{\mu}\psi\partial^{\mu}\psi - V(\phi,\chi,\psi),
\end{equation}
where $V(\phi,\chi,\psi)$ is the potential, that may contain self-interactions and interactions between the fields. It will be specified below, to give rise to first-order equations under the Bogomol'nyi procedure \cite{Bogomol'nyi}. Here we use natural units ($\hbar=1$ and $c=1$), dimensionless fields and space and time dimensions, with $x^\mu=(x^0=t,x^1=x)$ and $\partial_\mu\phi=(\partial\phi/\partial t=\dot\phi,\;
\partial\phi/ \partial x= \phi^\prime)$. The methodology follows the general case, in which the equations of motion are
\begin{align}
\begin{split}
 \partial_{\mu}\PR{f(\chi,\psi)\partial^{\mu}\phi} + \frac{\partial V}{\partial\phi} &= 0,
    \\    \partial_{\mu}\PR{g(\psi)\partial^{\mu}\chi} + \frac{\partial V}{\partial\chi} &= \frac{1}{2}\frac{\partial}{\partial\chi}\PR{f(\chi,\psi)}\partial_{\mu}\phi\partial^{\mu}\phi,
\\\partial_{\mu}\partial^{\mu}\psi + \frac{\partial V}{\partial\psi} &= \frac{1}{2}\frac{d}{d\psi}\PR{g(\psi)}\partial_{\mu}\chi\partial^{\mu}\chi + \frac{1}{2}\frac{\partial}{\partial\psi}\PR{f(\chi,\psi)}\partial_{\mu}\phi\partial^{\mu}\phi.
\end{split}
\end{align}
We notice that the scalar fields are, in principle, coupled by the potential $V(\phi,\chi,\psi)$ and by the functions $f(\chi,\psi)$ and $g(\psi)$. Since we will be searching for spatially localized structures described by static configurations, the above equations become
\begin{align}
\begin{split}
\label{Second_Order}
    \frac{d}{dx}\PR{f(\chi,\psi)\phi'} &= V_{\phi},
    \\
    \frac{d}{dx}\PR{g(\psi)\chi'} &= V_{\chi} + \frac{1}{2}\frac{\partial}{\partial\chi}\PR{f(\chi,\psi)}(\phi')^2,
    \\
    \psi'' &= V_{\psi} + \frac{1}{2}\frac{d}{d\psi}\PR{g(\psi)}(\chi')^2 + \frac{1}{2}\frac{\partial}{\partial\psi}\PR{f(\chi,\psi)}(\phi')^2,
\end{split}
\end{align}
where $\partial V/\partial \phi = V_{\phi}$, $\partial V/\partial \chi = V_{\chi}$ and $\partial V/\partial \psi = V_{\psi}$. The energy momentum tensor has the general form
 \begin{equation}
     T^{\mu\nu} = f(\chi,\psi)\partial^{\mu}\phi\partial^{\nu}\phi + g(\psi)\partial^{\mu}\chi\partial^{\nu}\chi + \partial^{\mu}\psi\partial^{\nu}\psi -\eta^{\mu\nu}\mathcal{L}.
 \end{equation}
It appears from translation invariance of the Lagrange density, and the corresponding energy density $T^{00}$ and stress $T^{11}$ are explicitly given by, for static fields,

 \begin{align}
\begin{split}
    T^{00} &= \frac{1}{2}f(\chi,\psi)\PR{\phi'}^2 + \frac{1}{2}g(\psi)\PR{\chi'}^2 + \frac{1}{2}\PR{\psi'}^2 + V(\phi,\chi,\psi),
    \\
    T^{11} &=  \frac{1}{2}f(\chi,\psi)\PR{\phi'}^2 + \frac{1}{2}g(\psi)\PR{\chi'}^2 + \frac{1}{2}\PR{\psi'}^2 - V(\phi,\chi,\psi).\label{stress}
\end{split}
\end{align}

Since the functions $f(\chi,\psi)$ and $g(\psi)$ appear linearly on the expression of the energy density, it is fundamental to consider these functions as nonnegative, to avoid configurations with negative energy contribution. Following the procedure developed by Bogomol'nyi \cite{Bogomol'nyi}, in which we investigate conditions that imply the presence of first-order differential equations and minimum energy, it is possible to rewrite the energy density as
\begin{eqnarray}
    T^{00} &=&\frac{f(\chi,\psi)}{2}\PR{\phi'\mp \frac{W_{\phi}}{f(\chi,\psi)}}^2 + \frac{g(\psi)}{2}\PR{\chi'\mp \frac{W_{\chi}}{g(\psi)}}^2 + \frac{1}{2}\PR{\psi'\mp W_{\psi}}^2+\nonumber\\
&&\qquad\qquad\qquad\qquad+\PR{ V(\phi,\chi,\psi) - \frac{1}{2}\frac{W_{\phi}^2}{f(\chi,\psi)} - \frac{1}{2}\frac{W_{\chi}^2}{g(\psi)} - \frac{1}{2}W_{\psi}^2} \pm \frac{dW}{dx},
\end{eqnarray}
where $W(\phi,\chi,\psi)$ is an auxiliary function that help us describe the methodology. So, if the potential is defined to be
\begin{equation}
\label{Potential}
V(\phi,\chi,\psi) = \frac{1}{2}\frac{W_{\phi}^2}{f(\chi,\psi)} + \frac{1}{2}\frac{W_{\chi}^2}{g(\psi)} + \frac{1}{2}W_{\psi}^2,
\end{equation}
and the first-order equations below are satisfied
\begin{align}
\begin{split}
\label{1oe}
    \phi' &= \pm\frac{W_{\phi}}{f(\chi,\psi)},
    \\
    \chi' &= \pm\frac{W_{\chi}}{g(\psi)},
    \\
    \psi' &= \pm W_{\psi},
\end{split}
\end{align}
it is then ensured that the system supports configurations with the energy minimized to 
\begin{equation}
    E_{B} = \int_{-\infty}^{+\infty}\!\!\!dx\; T^{00} = \int_{-\infty}^{+\infty}\!\!\!dx\;\frac{dW}{dx} .
\end{equation}
In this scenario, we can evaluate the energy of the solutions even without knowing the field configurations explicitly, because it only depends on the asymptotic values of $W(\phi,\chi,\psi)$, in the form
\begin{equation}
    E_{B} = \vert W(\phi(x\rightarrow +\infty), \chi(x\rightarrow +\infty), \psi(x\rightarrow +\infty)) - W(\phi(x\rightarrow -\infty), \chi(x\rightarrow -\infty), \psi(x\rightarrow -\infty)) \vert.
\end{equation}
It is important to notice that by writing the potential as in \eqref{Potential}, we are also fulfilling the stressless condition, that is, we are making $T^{11} = 0$ in Eq. \eqref{stress}. This result is also in accordance with Derrick's Theorem \cite{Der} and the Poincaré stability criterion \cite{Sta}. Moreover, we have used the first-order equations \eqref{1oe} to show that they obey the equations of motion \eqref{Second_Order} when $W$ satisfies  $W_{\phi\chi}=W_{\chi\phi}$, $W_{\chi\psi}=W_{\psi\chi}$ and $W_{\psi\phi}=W_{\phi\psi}$.

As we can see, to define each model we have to specify $W=W(\phi,\chi,\psi)$, $f(\chi,\psi)$ and $g(\psi)$. In this work we shall deal with several distinct possibilities; for simplicity, however, we always consider $W(\phi,\chi,\psi)=W_1(\phi) +W_2(\chi)+W_3(\psi)$, that is, $W$ is separable in the sum of three functions, each one depending on just one of three fields. This is of current interest, and we see that the third of the first-order equations \eqref{1oe}, involving the field $\psi$, is now independent of the other fields, so it can be solved directly, and  used to constrain the other fields. When working with the above first-order equations, we consider the upper signal, for simplicity. Moreover, in Sec. \ref{III} we deal with $f(\chi,\psi)=f(\chi)$, and in Sec. \ref{IV}
we take the more general possibility, with $f(\chi,\psi)$ depending on the two fields, $\chi$ and $\psi$. To define the potentials, we introduce two distinct parameters, $\alpha$ and $\beta$, real and positive, which are used to control the interaction of the fields and the shape
of the solutions. To depict the figures, we take some specific values for $\alpha$ and $\beta$, to show how the localized structures behave in terms of $\alpha$ and $\beta$.

Before we start the main investigation, let us briefly review the case of kinks in models described by a single real scalar field. This can be obtained by appropriately taking $\phi$ and $\chi$ as uniform configurations such that the above three-field model reduces to describe the $\psi$ field alone, with the first-order equation
$\psi^\prime = W_\psi$. We consider the $\psi^4$ model with $W_\psi=(1-\psi^2)$, which has the solution $\psi(x)=\tanh(x)$. In this work we consider the center of the solution at the origin, for simplicity. We can also consider the $\psi^6$ model, taking $W_\psi=\psi(1-\psi^2)$. In this case one of the solutions is $\psi(x)=
\sqrt{(1+\tanh(x))/2}$.
 In the figure below we depict these two solutions, which represent kinks of the standard $\psi^4$ and $\psi^6$ models, respectively. We see that the first solution vanishes at $x=0$, but the second one only vanishes asymptotically, in the limit $x\to-\infty$. As it was already emphasized in \cite{Const}, the zeroes of the solutions may be important to induce changes in the localized structures that can be used in applications of current interest. We further comment on this below.

\begin{figure}%
    \centering
    \centering{{\includegraphics[width=7.6cm]{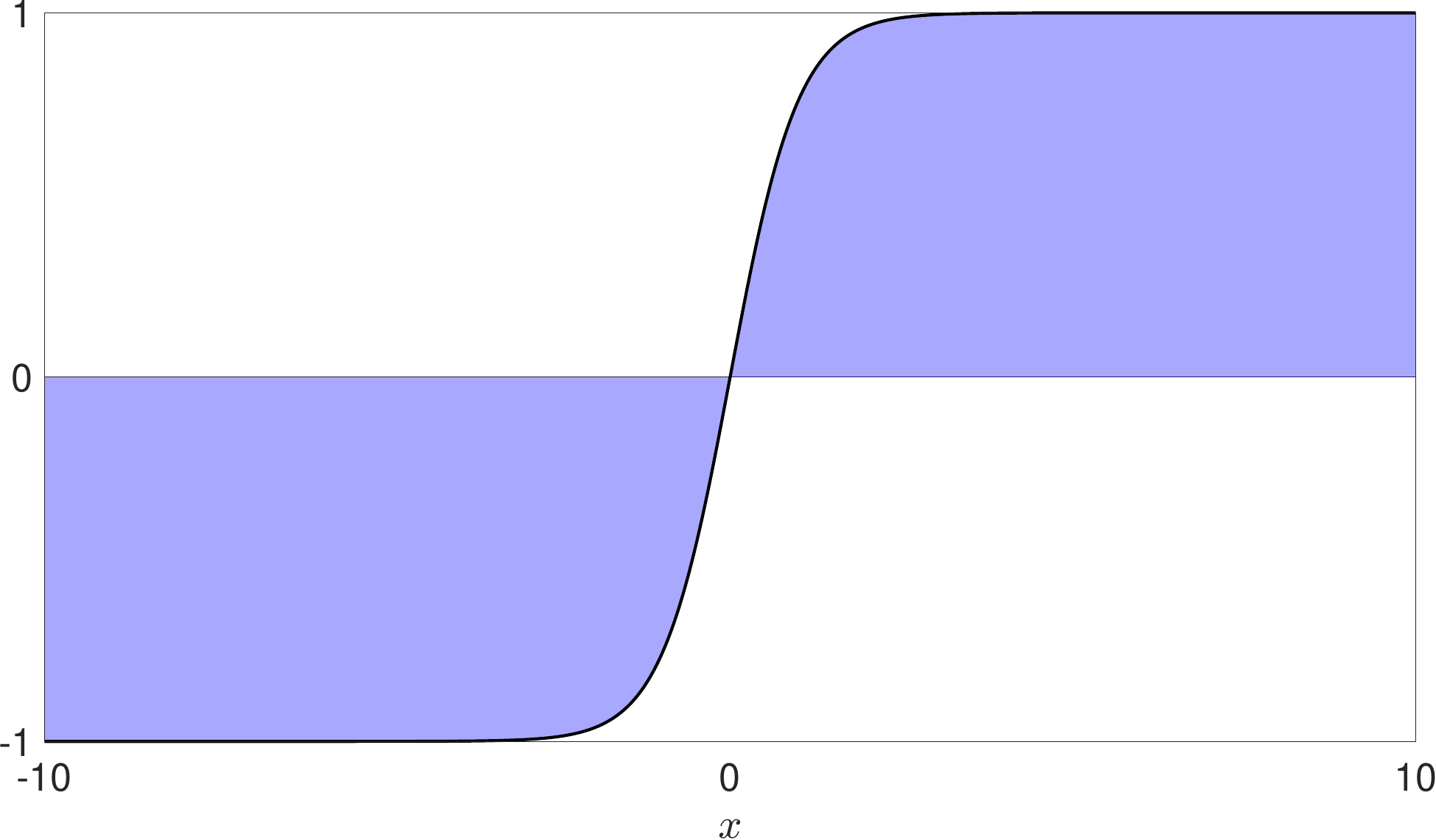} }}%
    \quad\quad\quad
\centering{{\includegraphics[width=7.6cm]{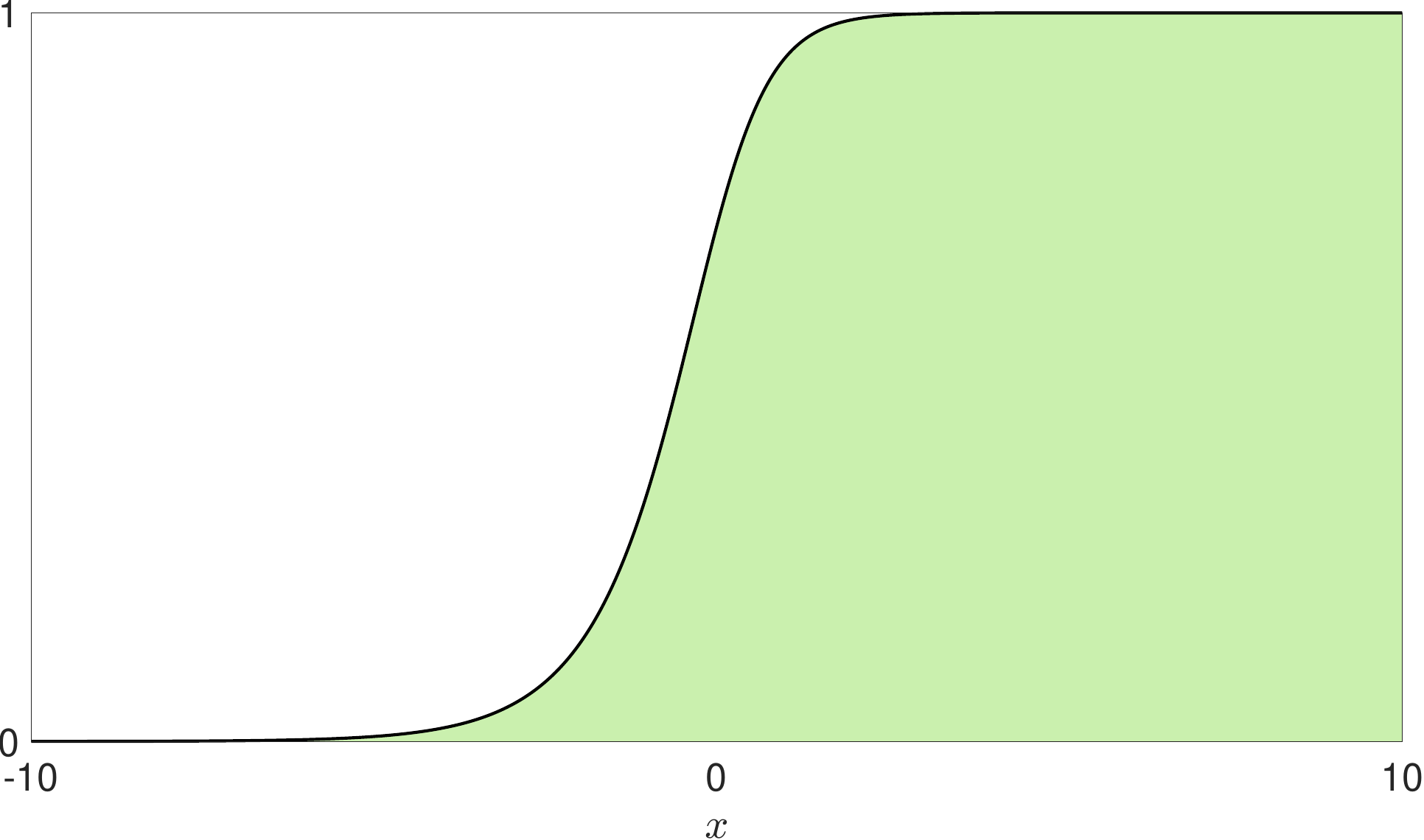} }}
\caption{Kinks of the standard $\psi^4$ (left) and $\psi^6$ (right) models. }\label{fig.1}
\end{figure}

\section{A Family of Models}\label{III}

To illustrate the procedure developed in Sec. \ref{II}, and analyze how the geometric constriction modifies the third field, we will primarily deal with a simpler version of the problem. This simplification occurs when we consider $f(\chi,\psi) = f(\chi)$, which will be studied below for several different models. 

\subsection{The 4-4-4 model}

Let us first consider the auxiliary function in the form
\begin{equation}
\label{W444}
W(\phi,\chi,\psi) = \phi - \frac{1}{3}\phi^3 + \alpha \chi - \frac{1}{3}\alpha\chi^3 + \beta\psi - \frac{1}{3}\beta \psi^3,
\end{equation}
where $\alpha$ and $\beta$ are positive real parameters. In this case, the three fields are of the fourth-order type, so we call the model the $4-4-4$ model. Since we still have the possibility to choose $f(\chi)$
and $g(\psi)$, below we consider several distinct cases.

\subsubsection{$f(\chi) = 1/\chi^2$\;\;\;and\:\;\;$g(\psi) = 1/\psi^2$} \label{IIIA1}
\begin{figure}%
    \centering
    \centering{{\includegraphics[width=7.6cm]{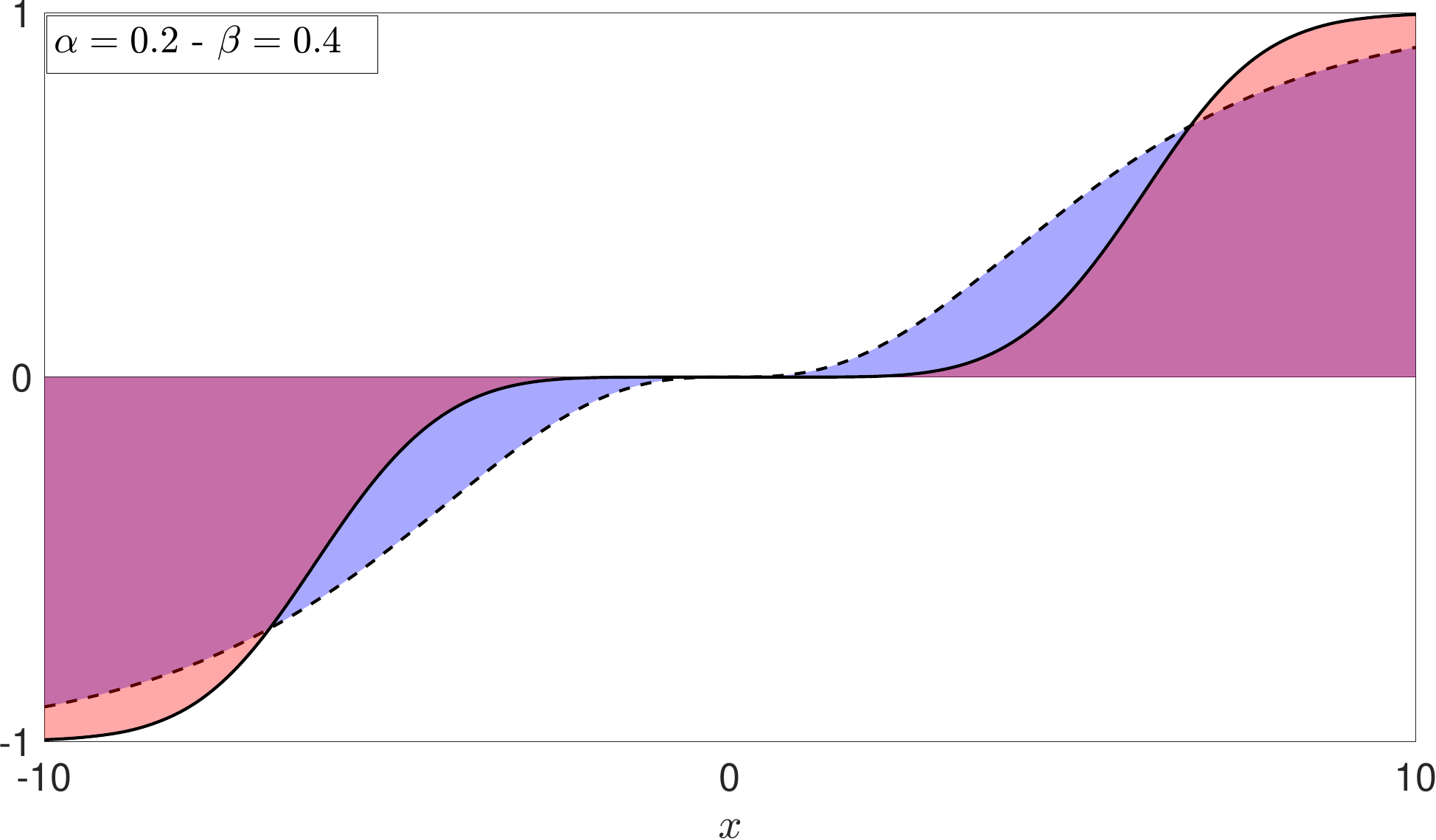} }}%
    \quad\quad\quad
\centering{{\includegraphics[width=7.6cm]{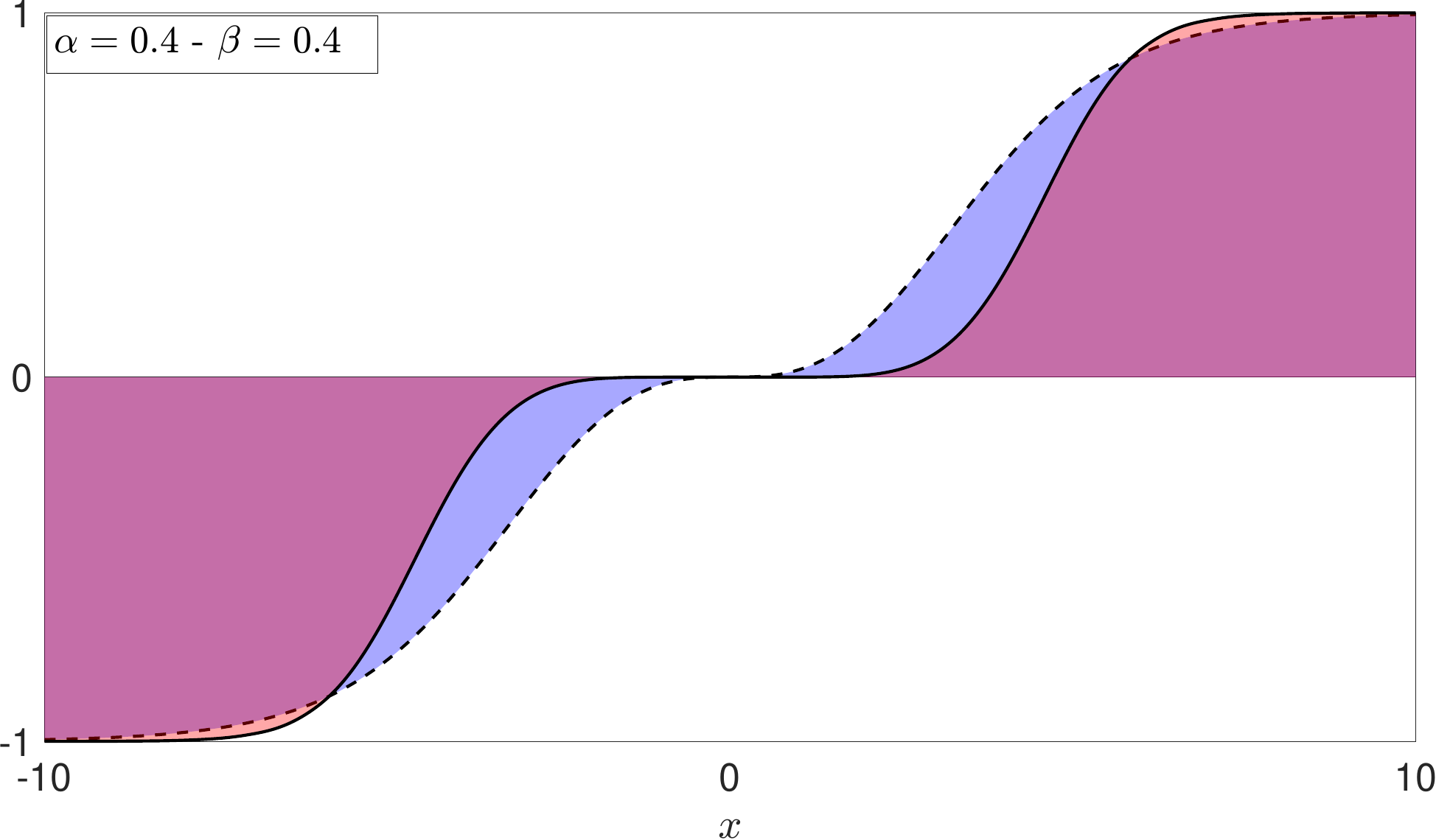} }}
\caption{The 4-4-4 model. Solutions $\phi$ (solid) and $\chi$ (dashed) for $f(\chi) = 1/\chi^2$ and $g(\psi) = 1/\psi^2$. }\label{fig.2}
\end{figure}

Here we have that $f(\chi) = 1/\chi^2$ and $g(\psi) = 1/\psi^2$, which are nonnegative functions, as required. Using Eq. \eqref{Potential}, the potential can be written as
\begin{equation}
\label{p444}
    V(\phi,\chi,\psi) = \frac{1}{2}\chi^2\PC{1-\phi^2}^2 + \frac{1}{2}\alpha^2\psi^2\PC{1-\chi^2}^2 + \frac{1}{2}\beta^2\PC{1-\psi^2}^2.
\end{equation}
This potential has minima at $\phi_{\pm} = \pm 1$, $\chi_{\pm} = \pm 1$ and $\psi_{\pm} = \pm 1$. In this model, the first-order equations \eqref{1oe} become
\begin{align}
\begin{split}
 \frac{d \psi}{dx} = \beta(1-\psi^2),
\;\;\;\;\;\;\;\;
\frac{d \chi}{dx} = \alpha\psi^2(1-\chi^2),
\;\;\;\;\;\;\;\;
\frac{d \phi}{dx} = \chi^2(1-\phi^2).
\end{split}
\end{align}
We notice that the first-order equation for $\psi$ can be solved independently, and has the solution
\begin{equation}
\label{psi4}
    \psi = \tanh(\beta x).
\end{equation}
Substituting this result in the first-order equation for the field $\chi$, we have that
\begin{equation}
\label{c444}
    \chi = \tanh\PC{\alpha Y_{\beta}(x)},
\end{equation}
where
\begin{equation}
\label{Y}
    Y_{\beta}(x) = x-\frac{1}{\beta}\tanh(\beta x),
\end{equation}
which represents the geometric constriction imposed by the field $\psi$ on the field $\chi$. Substituting Eq. \eqref{c444} in the first-order equation for the field $\phi$, we get
\begin{equation}
    \phi' = \tanh^2\PC{\alpha Y_{\beta}(x)}(1-\phi^2),
\end{equation}
which is much more complicated than the previous ones. Due to this it was solved numerically, and some of the $\phi$ and $\chi$ solutions are depicted in Fig \ref{fig.2}.

Interestingly, both fields $\phi$ and $\chi$ exhibit a double kink profile, but the plateau at the center of the solution $\phi$ is larger when compared to $\chi$. The formation of these plateaux is related to divergences of the coupling functions $f(\chi)$ and $g(\psi)$, as outlined in \cite{Const}. Since the field $\phi$ is modified by a field which is obtained having a plateau at a value where $f(\chi)$ diverges, this contributes to enlarge the plateau of $\phi$. The result suggests that the effect of $\chi$ on $\phi$ is further enhanced by the effect of $\psi$ on $\chi$.

\subsubsection{$f(\chi) = 1/\chi^2$\;\;\;\;\;$g(\psi) = 1/\cos^2(n\pi\psi)$}
\begin{figure}%
    \centering
    \centering{{\includegraphics[width=7.6cm]{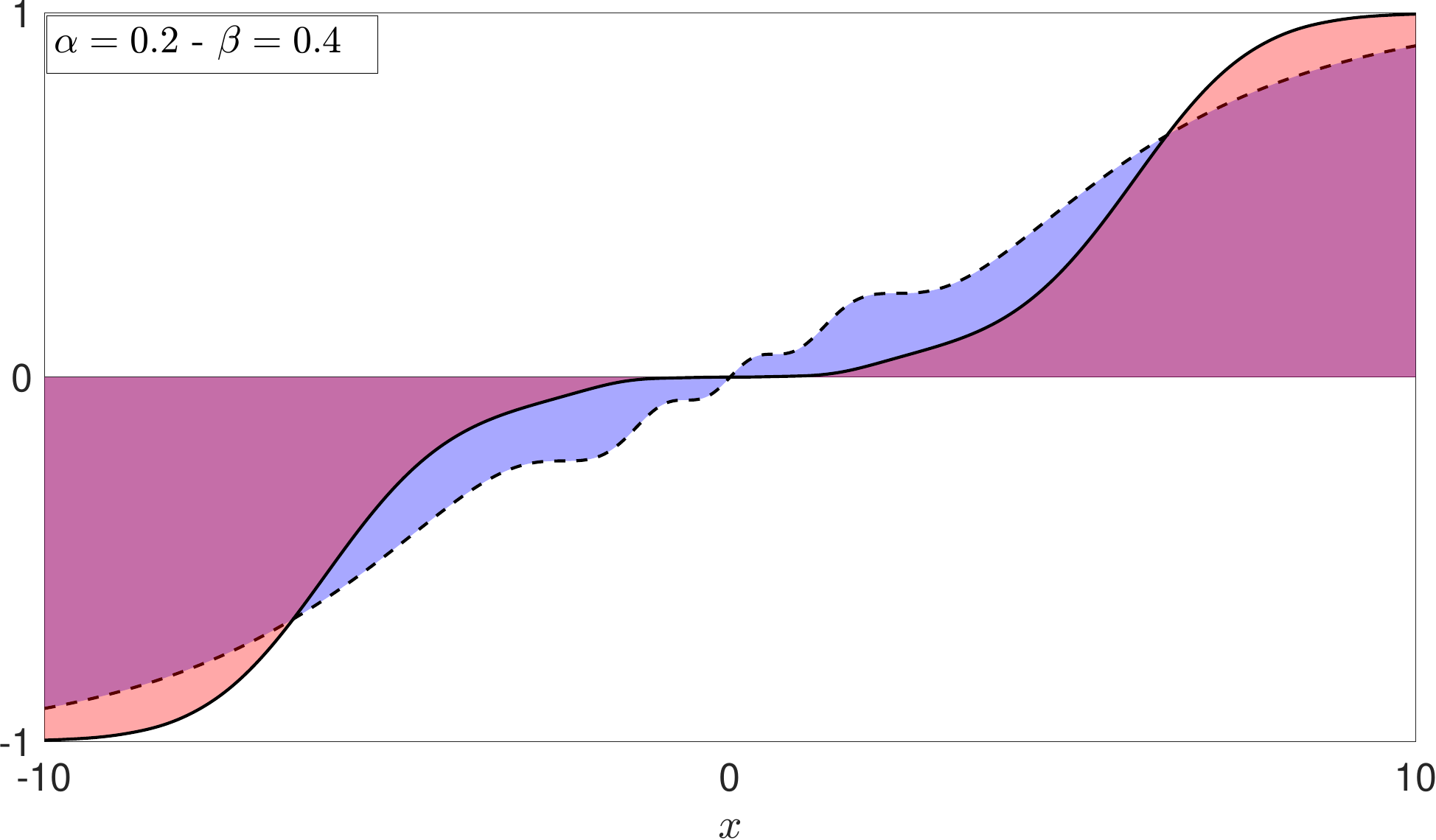} }}%
    \quad\quad\quad
    \centering{{\includegraphics[width=7.6cm]{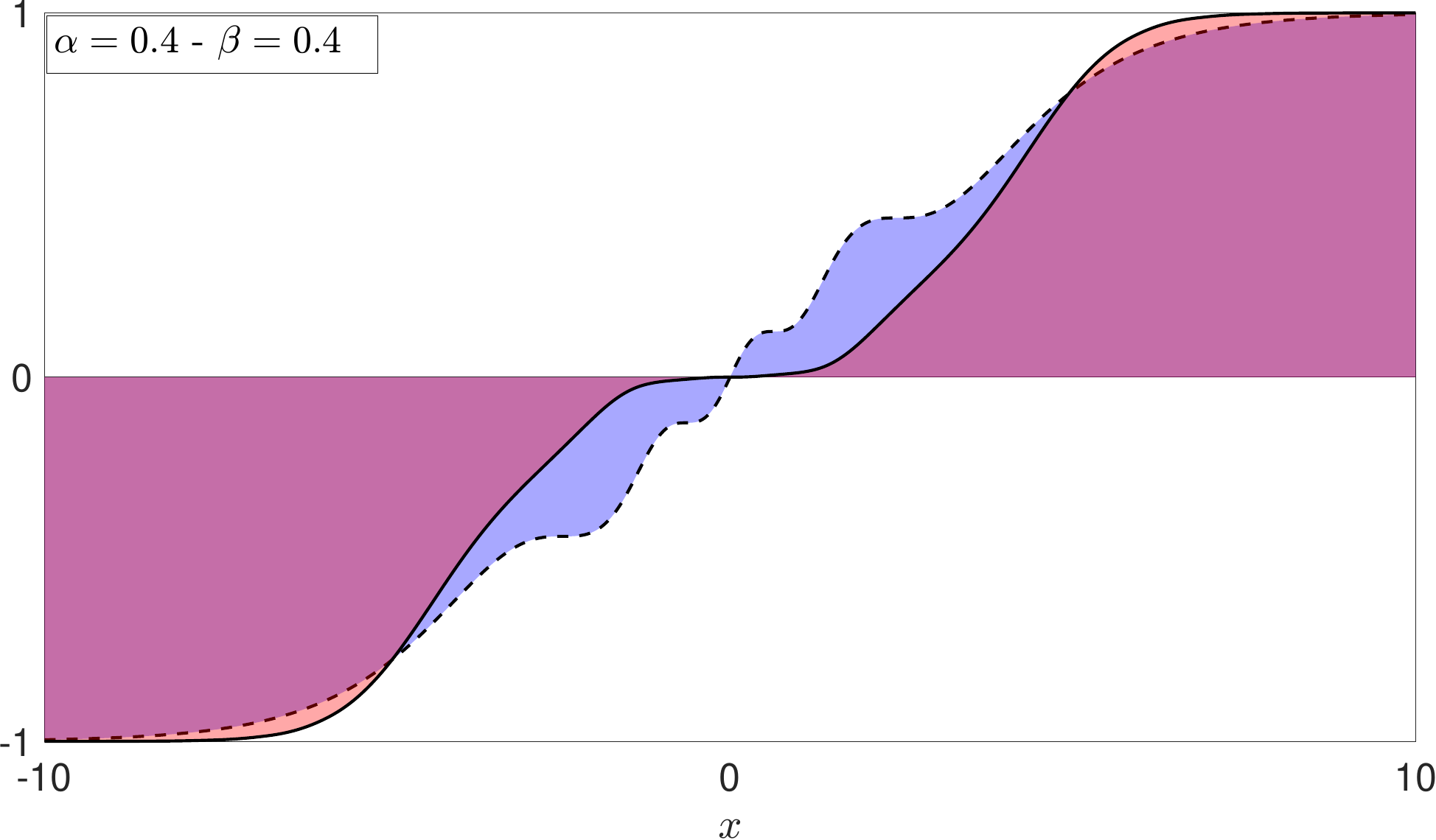} }}
\caption{The 4-4-4 model. Solutions $\phi$ (solid) and $\chi$ (dashed) for $f(\chi) = 1/\chi^2$, $g(\psi) = 1/\cos^2(n\pi\psi)$ and $n = 2$. }\label{fig.3}
\end{figure}

Another scenario to be investigated is determined by the functions $f(\chi) = 1/\chi^2$ and $g(\psi) = 1/\cos^2(n\pi\psi)$, $n \in \mathbb{N}$. Since we are still dealing with the 4-4-4 model, the auxiliary function $W(\phi,\chi,\psi)$ is defined as in the previous case \eqref{W444}. Using \eqref{Potential}, the potential has the form
\begin{equation}
    V(\phi,\chi,\psi) = \frac{1}{2}\chi^2\PC{1-\phi^2}^2 + \frac{1}{2}\alpha^2\cos^2(n\pi\psi)\PC{1-\chi^2}^2 + \frac{1}{2}\beta^2\PC{1-\psi^2}^2,
\end{equation}
which has the same set of minima as the previous case \eqref{p444}. In this model, the first-order equations \eqref{1oe} become
\begin{align}
\begin{split}
 \frac{d \psi}{dx} = \beta(1-\psi^2),
\;\;\;\;\;\;\;\;
\frac{d \chi}{dx} = \alpha\cos^2(n\pi\psi)(1-\chi^2),
\;\;\;\;\;\;\;\;
\frac{d \phi}{dx} = \chi^2(1-\phi^2).
\end{split}
\end{align} 
Considering that we have not changed the profile of the field $\psi$, its solution is given by \eqref{psi4}. Substituting this result in the first-order equation for the field $\chi$, we get as solution
\begin{equation}
    \chi = \tanh\PR{\alpha \PC{\frac{x}{2} + \frac{1}{4\beta}\PR{Ci(\xi_{n}^{+}(x)) - Ci(\xi_{n}^{-}(x))}}},
\end{equation}
where $Ci(z)$ is the cosine integral function, and $\xi_{n}^{\pm}(x) = 2n\pi(1\pm \tanh(\beta x))$. Substituting this result in the first-order equation for the field $\phi$, we get
\begin{equation}
    \phi' = \tanh^2\PR{\alpha \PC{\frac{x}{2} + \frac{1}{4\beta}\PR{Ci(\xi_{n}^{+}(x)) - Ci(\xi_{n}^{-}(x))}}}(1-\phi^2),
\end{equation}
which we solved numerically, giving the solution in Fig \ref{fig.3}.

We highlight the fact that, although the field $\chi$ has several plateaux, only one of them occurs when $\chi(x) = 0$, the value where $f(\chi)$ diverges. For this reason we have only one plateau on the field $\phi$. The other plateaux of $\chi$ slightly deform the $\phi$ solution outside its center.

\subsubsection{$f(\chi) = 1/\cos^2(n_{1}\pi\chi)$\;\;\;\;\;$g(\psi) = 1/\cos^2(n_{2}\pi\psi)$}
\begin{figure}[h!]
    \centering
    \centering{{\includegraphics[width=7.6cm]{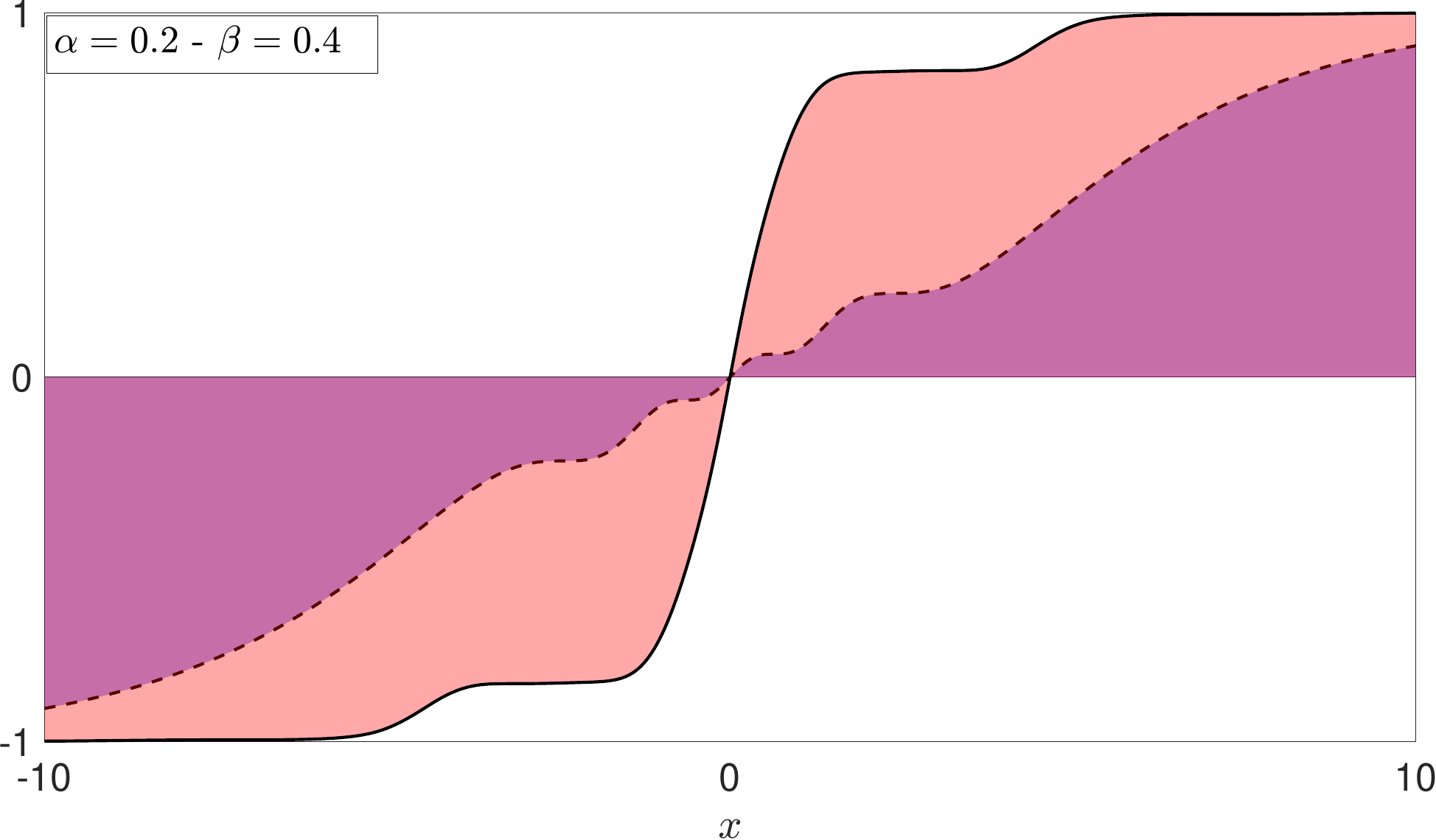} }}%
    \quad\quad\quad
    \centering{{\includegraphics[width=7.6cm]{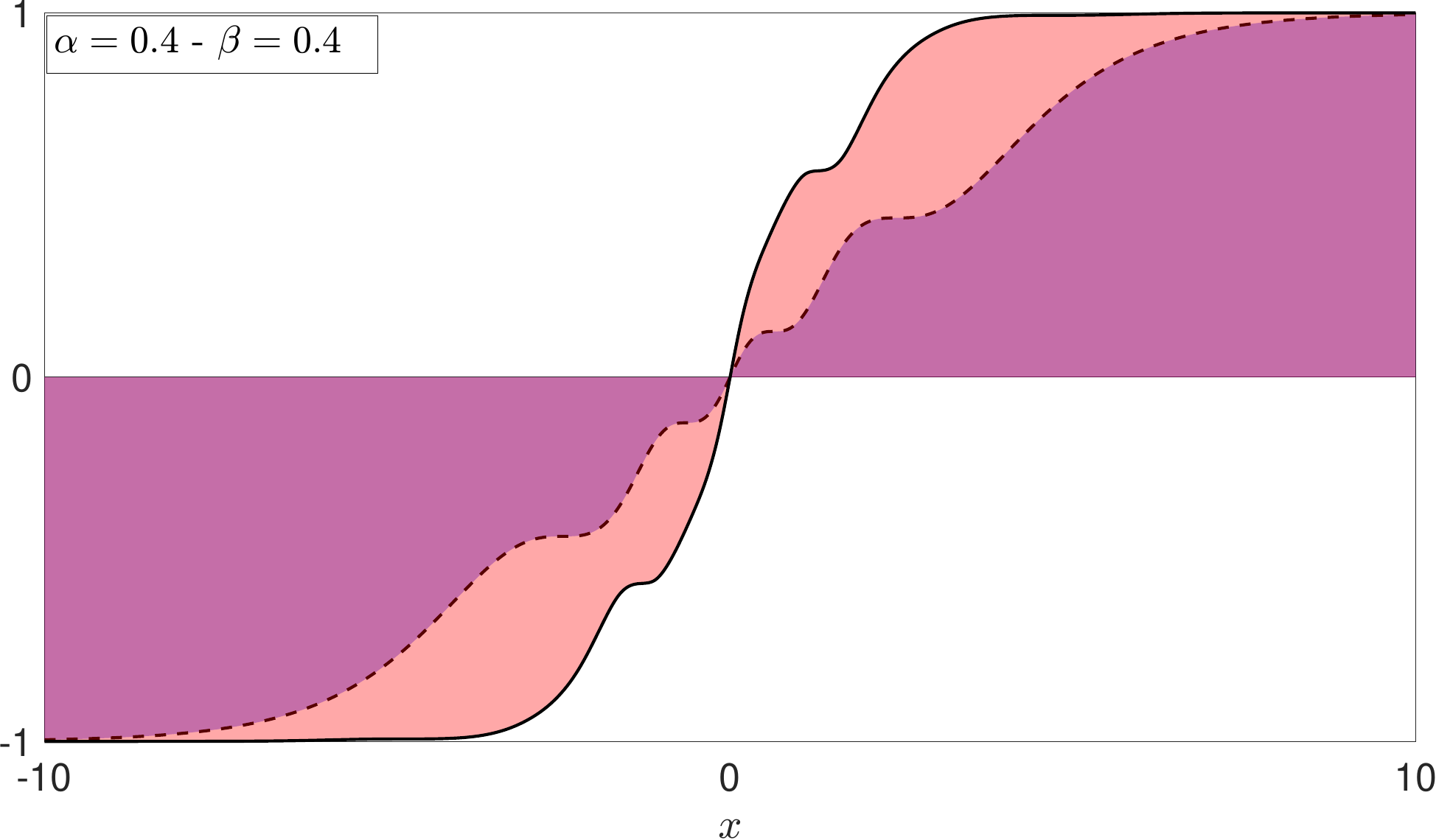} }}
\caption{The 4-4-4 model. Solutions $\phi$ (solid) and $\chi$ (dashed) for $f(\chi) = 1/\cos^2(n_{1}\pi\chi)$, $g(\psi) = 1/\cos^2(n_{2}\pi\psi)$ and $n_1 = n_2 = 2$. }\label{fig.4}
\end{figure}

Here we have that $f(\chi) = 1/\cos^2(n_{1}\pi\chi)$ and $g(\psi) = 1/\cos^2(n_{2}\pi\psi)$, where $n_{1}$,$ n_{2}\in \mathbb{N}$. Using \eqref{Potential}, the potential can be written as
\begin{equation}
    V(\phi,\chi,\psi) = \frac{1}{2}\cos^2(n_{1}\pi\chi)\PC{1-\phi^2}^2 + \frac{1}{2}\alpha^2\cos^2(n_{2}\pi\psi)\PC{1-\chi^2}^2 + \frac{1}{2}\beta^2\PC{1-\psi^2}^2,
\end{equation}
which has the same set of minima as the first case \eqref{p444}. In this model, the first-order equations \eqref{1oe} become
\begin{align}
\begin{split}
 \frac{d \psi}{dx} = \beta(1-\psi^2),
\;\;\;\;\;\;\;\;
\frac{d \chi}{dx} = \alpha\cos^2(n_{2}\pi\psi)(1-\chi^2),
\;\;\;\;\;\;\;\;
\frac{d \phi}{dx} = \cos^2(n_{1}\pi\chi)(1-\phi^2).
\end{split}
\end{align}
Since we have not changed the profile of the field $\psi$, its solution is given by \eqref{psi4}. Substituting this result in the first-order equation for the field $\chi$, we get as solution
\begin{equation}
    \chi = \tanh\PR{\alpha \PC{\frac{x}{2} + \frac{1}{4\beta}\PR{Ci(\xi_{n_{2}}^{+}(x)) - Ci(\xi_{n_{2}}^{-}(x))}}},
\end{equation}
where $Ci(z)$ is the already mentioned cosine integral function, and $\xi_{n_{2}}^{\pm}(x) = 2n_{2}\pi(1\pm \tanh(\beta x))$. Substituting this result in the first-order equation for the field $\phi$, we get
\begin{equation}
    \phi' = \cos^2\chav{n_{1}\pi\tanh\PR{\alpha \PC{\frac{x}{2} + \frac{1}{4\beta}\PR{Ci(\xi_{n_{2}}^{+}(x) - Ci(\xi_{n_{2}}^{-}(x))}}}}(1-\phi^2),
\end{equation}
which we solved numerically, giving the solutions depicted in Fig \ref{fig.4}.
We notice that the major modifications that appear in the field $\phi$ are very similar to the ones presented by $\chi$. Although $\chi(0) = 0$, we do not have the formation of a plateau around $\phi(0)$. This is a consequence of the fact that $f(\chi) = 1/\cos^2(n_{1}\pi\chi)$ does not diverge when $\chi$ goes to zero.\\

\subsection{Beyond the 4-4-4 model}

Above, we have shown how the definition of the functions $f(\chi)$ and $g(\psi)$ works to modify the solution of $\phi$, inducing internal structures and deforming its shape. Let us now investigate other possibilities, adding the sixth-order type of self-interaction for some of the fields. To accomplish that, we shall consider several distinct possibilities.

\subsubsection{The 4-4-6 model}

\begin{figure}[t!]
    \centering
    \centering{{\includegraphics[width=7.6cm]{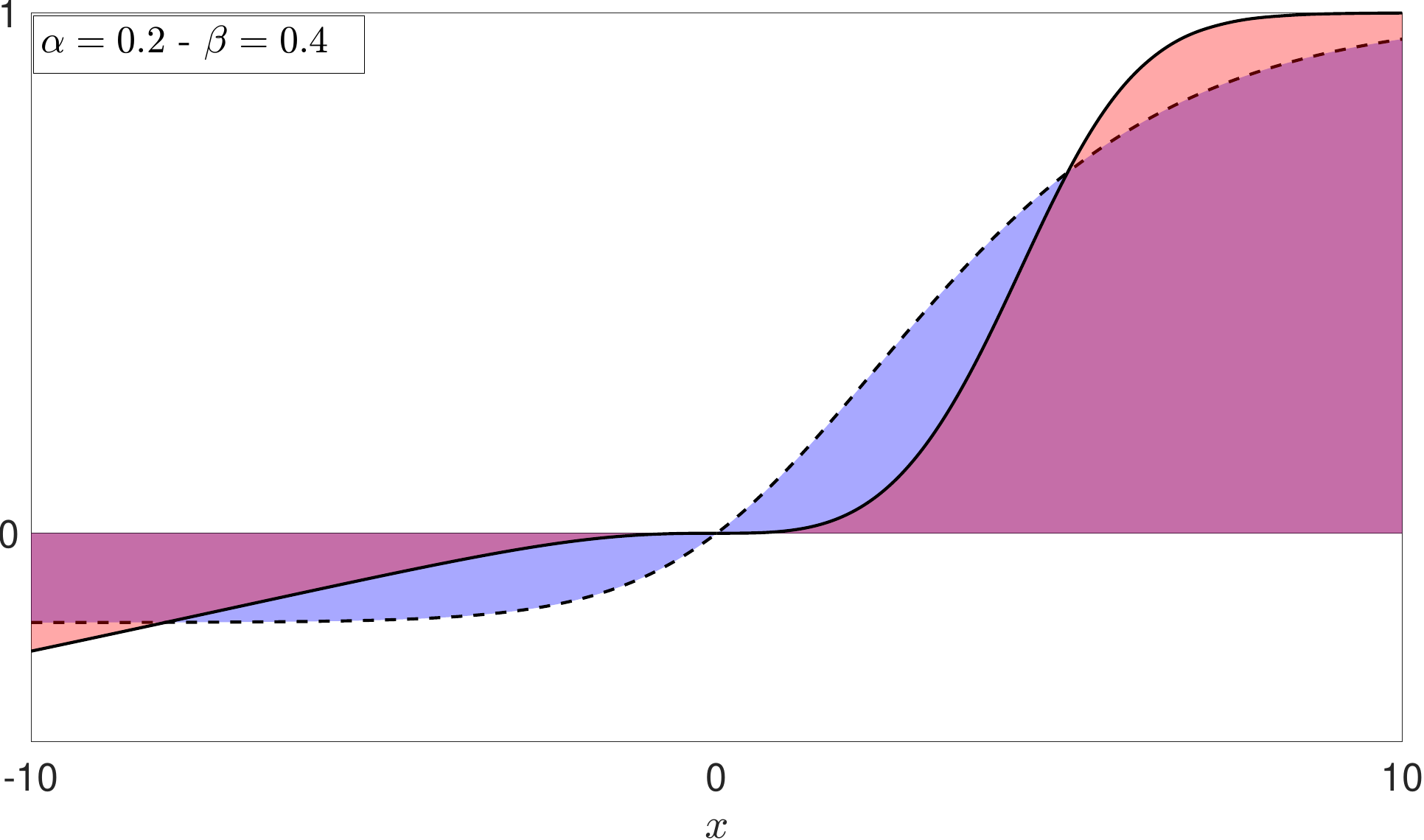} }}%
    \quad\quad\quad
    \centering{{\includegraphics[width=7.6cm]{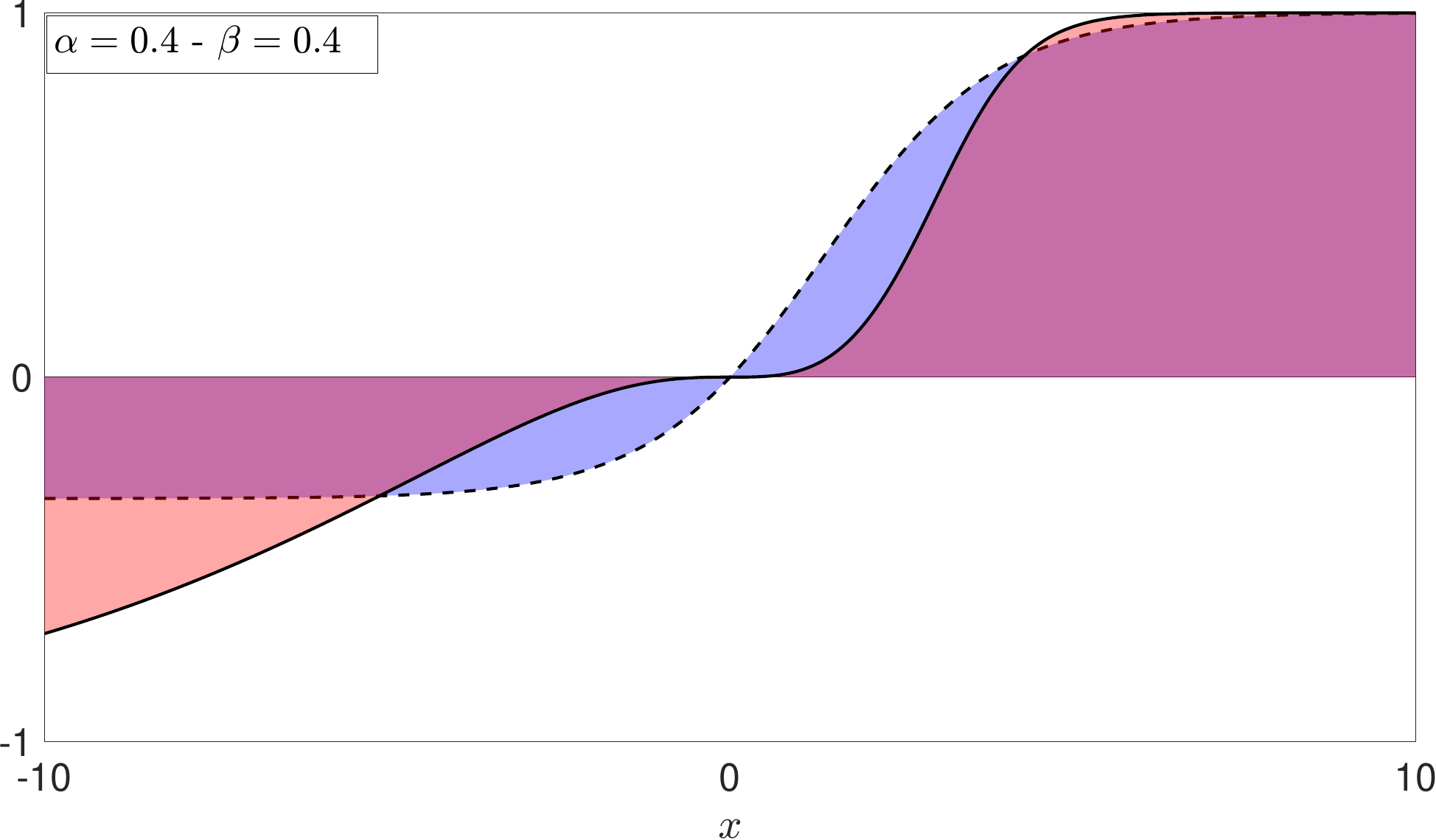} }}
\caption{The 4-4-6 model. Solutions $\phi$ (solid) and $\chi$ (dashed) for $f(\chi) = 1/\chi^2$ and $g(\psi) = 1/\psi^2$. }\label{fig.5}
\end{figure}

We have that $f(\chi) = 1/\chi^2$ and $g(\psi) = 1/\psi^2$. Here, however, we take the field $\psi$ to be governed by the $\psi^6$ potential. The auxiliary function is then defined as
\begin{equation}
W(\phi,\chi,\psi) = \phi - \frac{1}{3}\phi^3 +  \alpha\chi - \frac{1}{3}\alpha\chi^3 + \frac{1}{2}\beta\psi^2 - \frac{1}{4}\beta \psi^4.
\end{equation}
 Using \eqref{Potential}, the potential can be written as
\begin{equation}
V(\phi,\chi,\psi) = \frac{1}{2}\chi^2\PC{1-\phi^2}^2 + \frac{1}{2}\alpha^2\psi^2\PC{1-\chi^2}^2 + \frac{1}{2}\beta^2\psi^2\PC{1-\psi^2}^2.
\end{equation}
This potential has minima at $\phi_{\pm} = \pm 1$, $\chi_{\pm} = \pm 1$ and $\psi_{\pm} = \pm 1$. Besides this, it also has some continuum lines of minima: at $\psi = \chi = 0$, and $\phi$ arbitrary; and at $\psi = 0$, $\phi = \pm 1$ and $\chi$ arbitrary. In this model, the first-order equations \eqref{1oe} become
\begin{align}
\begin{split}
 \frac{d \psi}{dx} = \beta\psi(1-\psi^2),
\;\;\;\;\;\;\;\;
\frac{d \chi}{dx} = \alpha\psi^2(1-\chi^2),
\;\;\;\;\;\;\;\;
\frac{d \phi}{dx} = \chi^2(1-\phi^2).
\end{split}
\end{align}
Since the equation for $\psi$ is independent from the other fields, it can be easily solved to give
\begin{equation}
\label{psi6}
    \psi = \sqrt{\frac{1}{2}\PC{1+\tanh(\beta x)}}.
\end{equation}
Here we are taking the plus signal inside the square root. Substituting this result in the first-order equation for the field $\chi$, we have as solution
\begin{equation}
    \chi = \tanh\PC{ \alpha Z_{\beta}(x)},
\end{equation}
where
\begin{equation}
\label{Z}
    Z_{\beta}(x) = \frac{x}{2} +\frac{1}{2\beta}\ln(\cosh(\beta x)),
\end{equation}
which is the geometrical constriction imposed to the model by the field $\psi$ and the function $g(\psi)$. Substituting this result in the first-order equation for the field $\phi$, we get
\begin{equation}
    \phi' = \tanh^2\PC{\alpha Z_{\beta}(x)}(1-\phi^2),
\end{equation}
which we solved numerically, giving the solutions depicted in Fig \ref{fig.5}. We notice that, although $\phi$ exhibit an asymmetric behaviour, due to the constriction induced by the asymmetric solution of the field $\chi$, it still asymptotically goes to $\pm 1$. Also, since $\chi(0) = 0$, a plateau appears at $\phi(0)$.

\subsubsection{The 4-6-4 model}
We have that $f(\chi) = 1/\chi^2$ and $g(\psi) = 1/\psi^2$, but now the auxiliary function is defined as
\begin{equation}
W(\phi,\chi,\psi) = \phi - \frac{1}{3}\phi^3 +  \frac{1}{2}\alpha\chi^2 - \frac{1}{4}\alpha\chi^4 + \beta\psi - \frac{1}{3}\beta \psi^3.
\end{equation}
Using \eqref{Potential}, the potential can be written as
\begin{equation}
    V(\phi,\chi,\psi) = \frac{1}{2}\chi^2\PC{1-\phi^2}^2 + \frac{1}{2}\alpha^2\psi^2\chi^2\PC{1-\chi^2}^2 + \frac{1}{2}\beta^2\PC{1-\psi^2}^2.
\end{equation}

\begin{figure}[h!]
    \centering
    \centering{{\includegraphics[width=7.6cm]{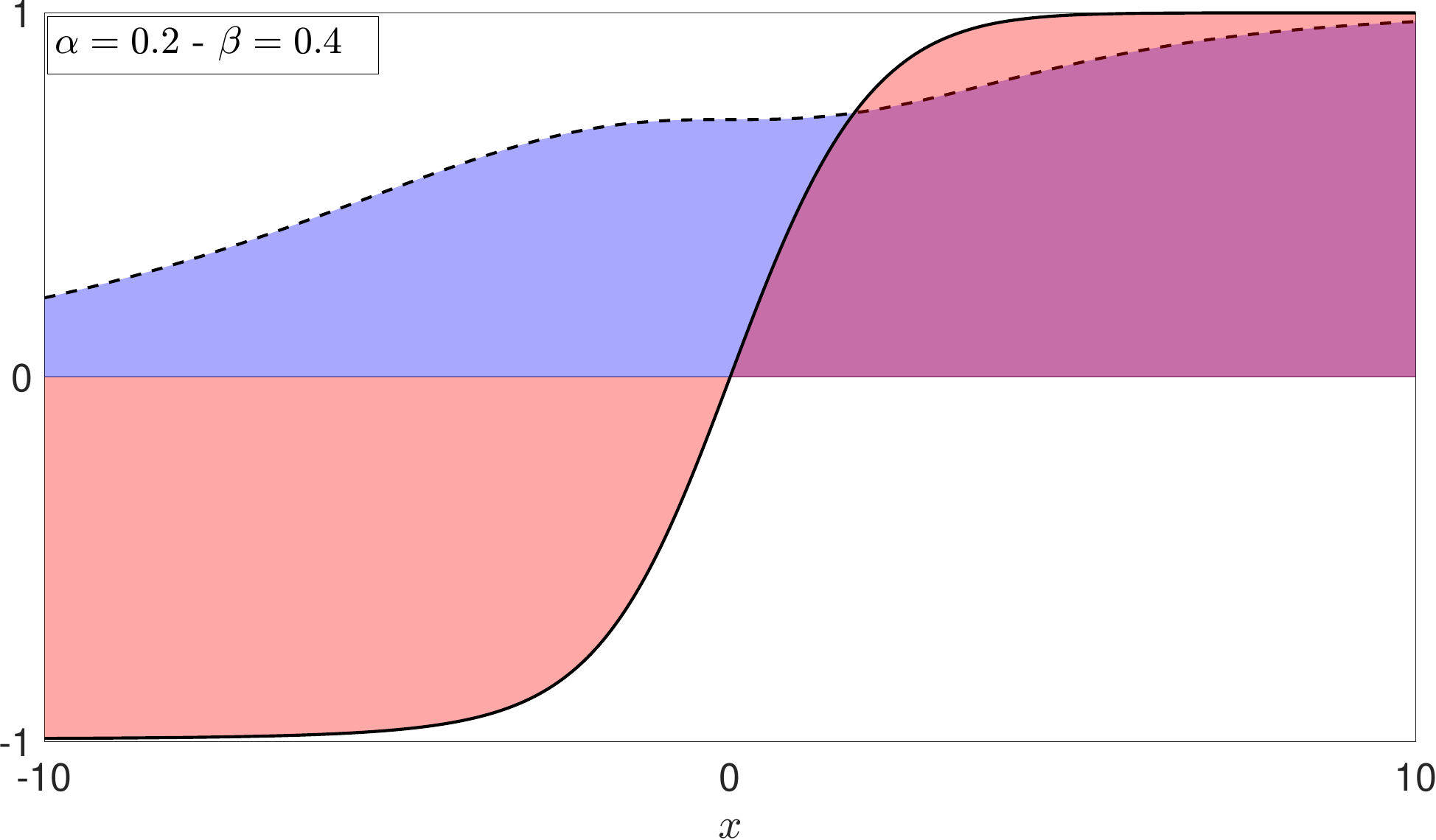} }}%
    \quad\quad\quad
    \centering{{\includegraphics[width=7.6cm]{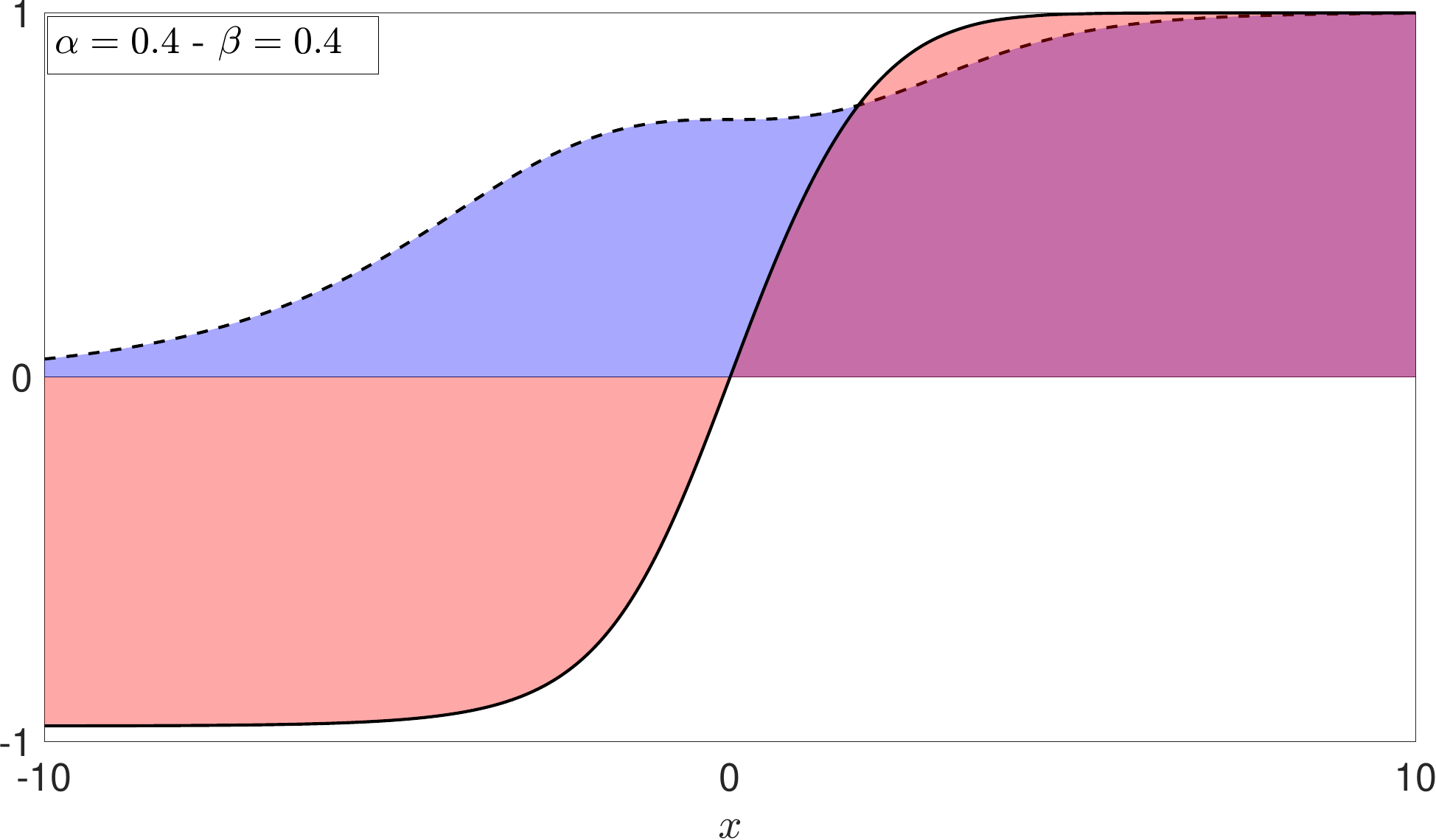} }}
\caption{The 4-6-4 model. Solutions $\phi$ (solid) and $\chi$ (dashed) for $f(\chi) = 1/\chi^2$ and $g(\psi) = 1/\psi^2$. }\label{fig.6}
\end{figure}

This potential has minima at $\phi_{\pm} = \pm 1$, $\chi_{\pm} = \pm 1$ and $\psi_{\pm} = \pm 1$. Besides this, it also has continuum lines of minima at $\psi = \pm 1$, $\chi = 0$ and $\phi$ arbitrary. In this model, the first-order equations \eqref{1oe} become

\begin{align}
\begin{split}
 \frac{d \psi}{dx} = \beta(1-\psi^2),
\;\;\;\;\;\;\;\;
\frac{d \chi}{dx} = \alpha\psi^2\chi(1-\chi^2),
\;\;\;\;\;\;\;\;
\frac{d \phi}{dx} = \chi^2(1-\phi^2).
\end{split}
\end{align}
Since we are dealing with a $\psi^4$ model, the solution was already calculated in Eq. \eqref{psi4}. Substituting this result in the first-order equation for the field $\chi$, we get as solution
\begin{equation}
    \chi = \sqrt{\frac{1}{2}\PR{1+\tanh\PC{\alpha Y_{\beta}(x)}}}.
\end{equation}
Here we are also taking the plus signal inside the square root. $Y_{\beta}(x)$ was already defined in Eq. \eqref{Y}. Substituting the previous result in the first-order equation for the field $\phi$, we get
\begin{equation}
    \phi' = \frac{1}{2}\PR{1+\tanh\PC{\alpha Y_{\beta}(x)}}(1-\phi^2),
\end{equation}
which we solved numerically, giving the solutions displayed in Fig \ref{fig.6}. We notice that, since $\chi$ goes to zero in the limit where $x\rightarrow -\infty$, it induces major modifications on the left tail of $\phi(x)$, and this also explains why there is no plateau in the third field solution.

\subsubsection{The 4-6-6 model}
We have that $f(\chi) = 1/\chi^2$ and $g(\psi) = 1/\psi^2$. The auxiliary function is defined as

\begin{equation}
    W(\phi,\chi,\psi) = \phi - \frac{1}{3}\phi^3 +  \frac{1}{2}\alpha\chi^2 - \frac{1}{4}\alpha\chi^4 + \frac{1}{2}\beta\psi^2 - \frac{1}{4}\beta \psi^4.
\end{equation}
Using \eqref{Potential}, the potential is
\begin{equation}
    V(\phi,\chi,\psi) = \frac{1}{2}\chi^2\PC{1-\phi^2}^2 + \frac{1}{2}\alpha^2\psi^2\chi^2\PC{1-\chi^2}^2 + \frac{1}{2}\beta^2\psi^2\PC{1-\psi^2}^2.
\end{equation}

\begin{figure}%
    \centering
    \centering{{\includegraphics[width=7.6cm]{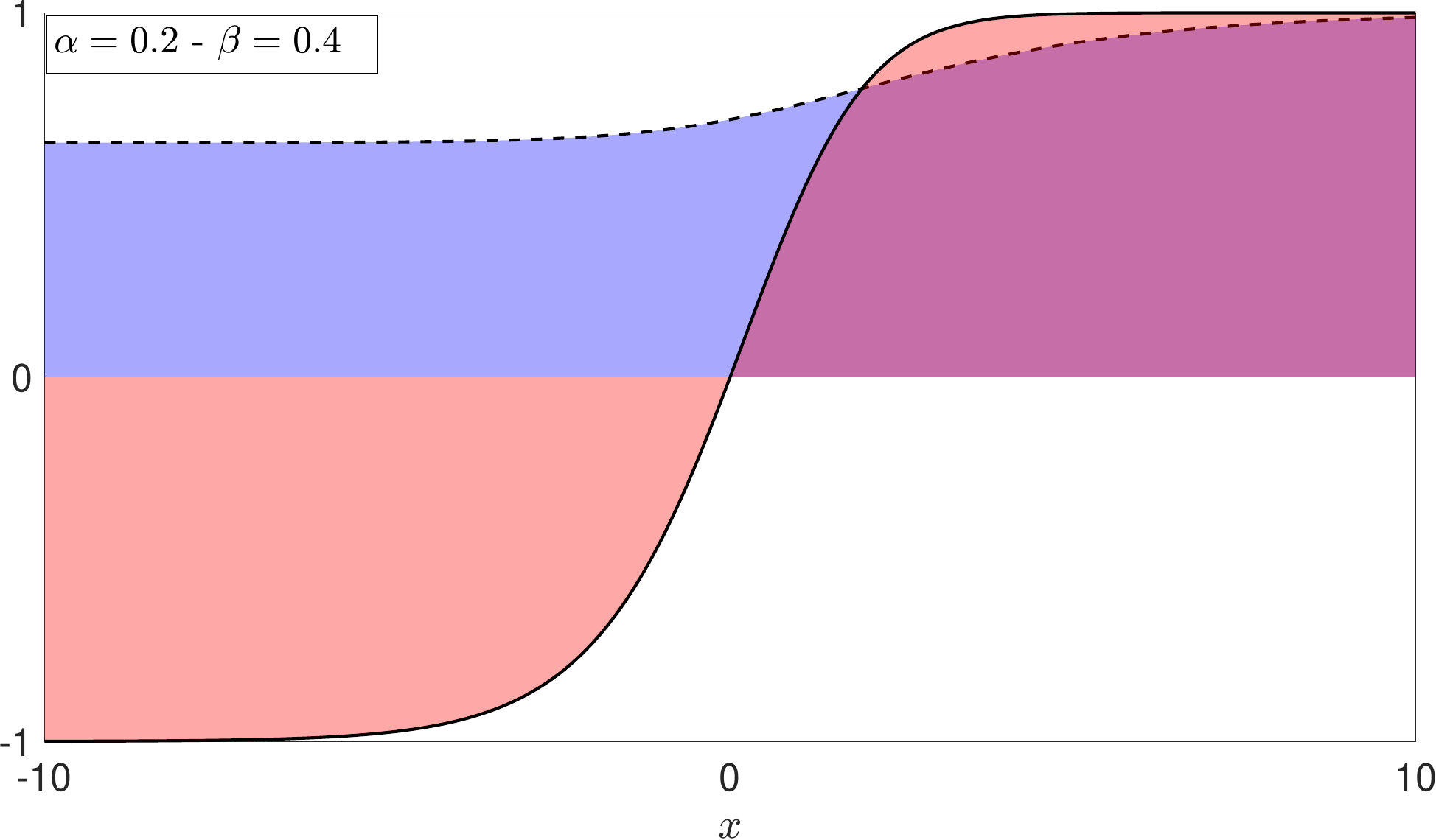} }}%
    \quad\quad\quad
    \centering{{\includegraphics[width=7.6cm]{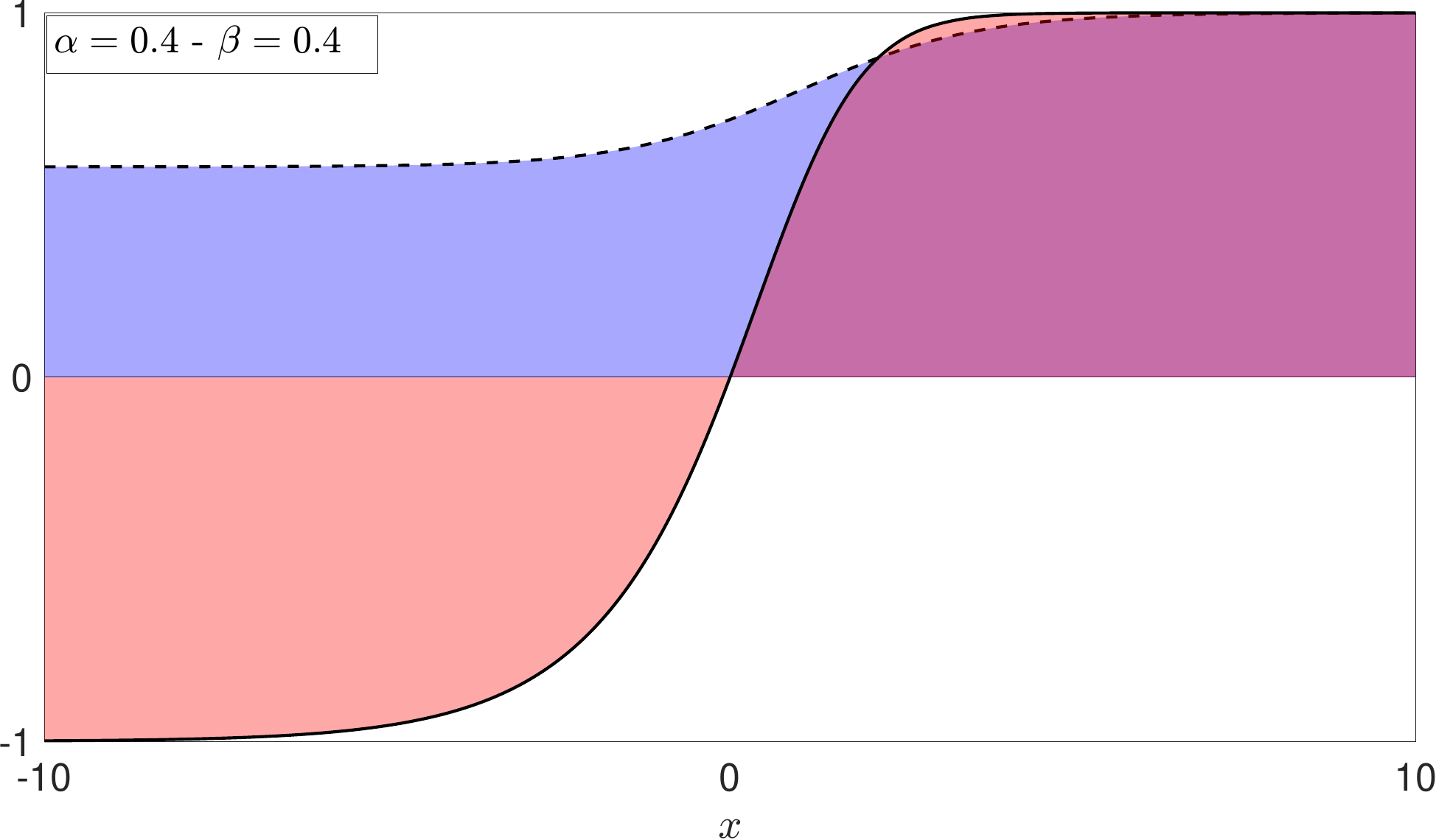} }}
\caption{The 4-6-6 model. Solutions $\phi$ (solid) and $\chi$ (dashed) for $f(\chi) = 1/\chi^2$ and $g(\psi) = 1/\psi^2$. }\label{fig.7}
\end{figure}

This potential has minima at $\phi_{\pm} = \pm 1$, $\chi_{\pm} = \pm 1$ and $\psi_{\pm} = \pm 1$. It also has several lines of continuum minima. In this model, the first-order equations \eqref{1oe} become

\begin{align}
\begin{split}
 \frac{d \psi}{dx} = \beta\psi(1-\psi^2),
\;\;\;\;\;\;\;\;
\frac{d \chi}{dx} = \alpha\psi^2\chi(1-\chi^2),
\;\;\;\;\;\;\;\;
\frac{d \phi}{dx} = \chi^2(1-\phi^2).
\end{split}
\end{align}
Since we are dealing with a $\psi^6$ model, the solution was already calculated in Eq. \eqref{psi6}. Substituting this result in the first-order equation for the field $\chi$, we get the solution
\begin{equation}
    \chi = \sqrt{\frac{1}{2}\PR{1+\tanh\PC{\alpha Z_{\beta}(x)}}}.
\end{equation}
Here we are also taking the plus signal inside the square root. $Z_{\beta}(x)$ was already defined in Eq. \eqref{Z}. Substituting this result in the first-order equation for the field $\phi$, we get
\begin{equation}
    \phi' = \frac{1}{2}\PR{1+\tanh\PC{\alpha Z_{\beta}(x)}}(1-\phi^2),
\end{equation}
which we solved numerically, giving the solutions displayed in Fig \ref{fig.7}. We highlight the fact that since $\chi(x)$ never reaches zero, then $f(\chi)$ never diverges, so it modifies neither the internal structures nor the asymptotic values of the field $\phi$. However, since $\chi(x)$ is asymmetric, it induces an asymmetric profile to the field $\phi$.

\section{Another Family of Models}\label{IV}

In this section we take $f(\chi,\psi)$ as the function that constrains the third field. Since we have already explored how different models affect the geometric constriction, here we will only consider the 4-4-4 model.

\subsection{$f(\chi,\psi) = 1/(\chi+\psi)^2$\;\;\;\;\;$g(\psi) = 1/\psi^2$}
We have that $f(\chi,\psi) = 1/(\chi+\psi)^2$ and $g(\psi) = 1/\psi^2$. The auxiliary function for this model was defined in Eq. \eqref{W444}. Using \eqref{Potential}, the potential can be written as
\begin{equation}
    V(\phi,\chi,\psi) = \frac{1}{2}(\chi+\psi)^2\PC{1-\phi^2}^2 + \frac{1}{2}\alpha^2\psi^2\PC{1-\chi^2}^2 + \frac{1}{2}\beta^2\PC{1-\psi^2}^2.
\end{equation}

\begin{figure}%
    \centering
    \centering{{\includegraphics[width=7.6cm]{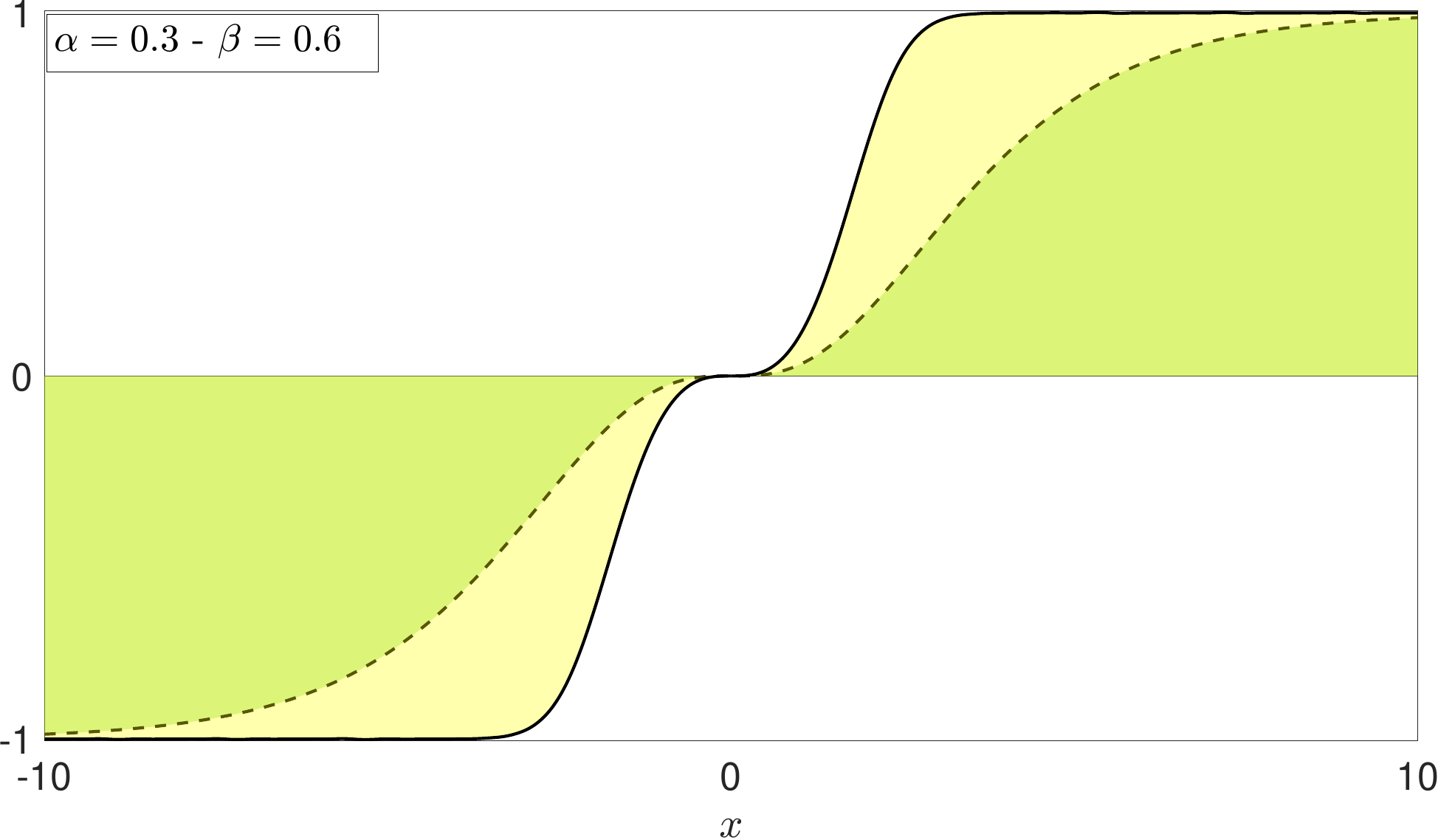} }}%
    \quad\quad\quad
    \centering{{\includegraphics[width=7.6cm]{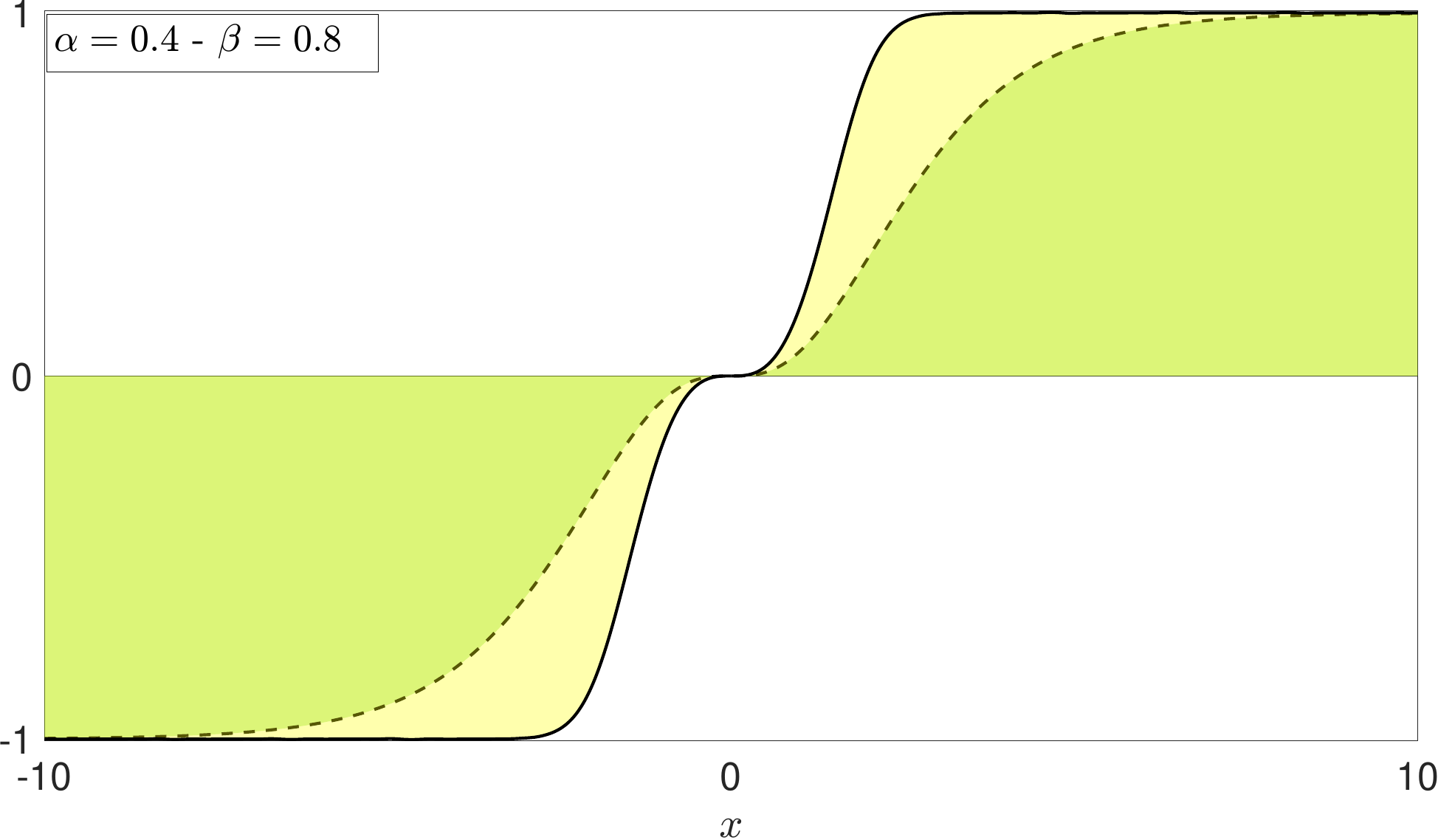} }}
\caption{The 4-4-4 model. Solutions $\phi$ (solid) and $\chi$ (dashed) for $f(\chi,\psi) = 1/(\chi+\psi)^2$ and $g(\psi) = 1/\psi^2$. }\label{fig.8}
\end{figure}

This potential has minima at $\phi_{\pm} = \pm 1$, $\chi_{\pm} = \pm 1$ and $\psi_{\pm} = \pm 1$. Besides this, it also has two other lines of continuum minima. In this model, the first-order equations \eqref{1oe} become
\begin{align}
\begin{split}
 \frac{d \psi}{dx} = \beta(1-\psi^2),
\;\;\;\;\;\;\;\;
\frac{d \chi}{dx} = \alpha\psi^2(1-\chi^2),
\;\;\;\;\;\;\;\;
\frac{d \phi}{dx} = (\psi+\chi)^2(1-\phi^2).
\end{split}
\end{align}
Since we are dealing with the $\psi^4$ model, the solution was already calculated in Eq. \eqref{psi4}. Substituting this result in the first-order equation for the field $\chi$, we get the solution
\begin{equation}\label{45}
    \chi = \tanh\PR{\alpha Y_{\beta}(x)}.
\end{equation}
Applying this result in the first-order equation for the field $\phi$, we get
\begin{equation}
    \phi' = \PR{\tanh\PC{\alpha Y_{\beta}(x)}+\tanh(\beta x)}^2(1-\phi^2),
\end{equation}
which we solved numerically, giving the solutions depicted in Fig \ref{fig.8}. Differently from the discussion made on the model described in Sec. \ref{IIIA1}, here the plateau of the third field is smaller when compared to $\chi(x)$. This behaviour arises because $\chi + \psi$ has no plateau at $x = 0$, so it modifies $\phi$ in a very similar way as in the standard $\chi^4$ case.

On the other hand, if we define $f(\chi, \psi) = 1/(\chi - \psi)^2$, a internal structure arises, with a behaviour which is similar to the above one. However, in this case the solution of the third field $\phi(x)$ also has its asymptotic values modified, due to the divergence that arises on $f(\chi,\psi)$ when $x \rightarrow \pm \infty$.

\subsection{$f(\chi,\psi) = 1/(\chi\psi)^2$\;\;\;\;\;$g(\psi) = 1/\psi^2$}
We have that $f(\chi,\psi) = 1/(\chi\psi)^2$ and $g(\psi) = 1/\psi^2$. Since we are dealing with the 4-4-4 model, the auxiliary function is defined as in Eq. \eqref{W444}. Using \eqref{Potential}, the potential reads
\begin{equation}
    V(\phi,\chi,\psi) = \frac{1}{2}\chi^2\psi^2\PC{1-\phi^2}^2 + \frac{1}{2}\alpha^2\psi^2\PC{1-\chi^2}^2 + \frac{1}{2}\beta^2\PC{1-\psi^2}^2.
\end{equation}

\begin{figure}%
    \centering
    \centering{{\includegraphics[width=7.6cm]{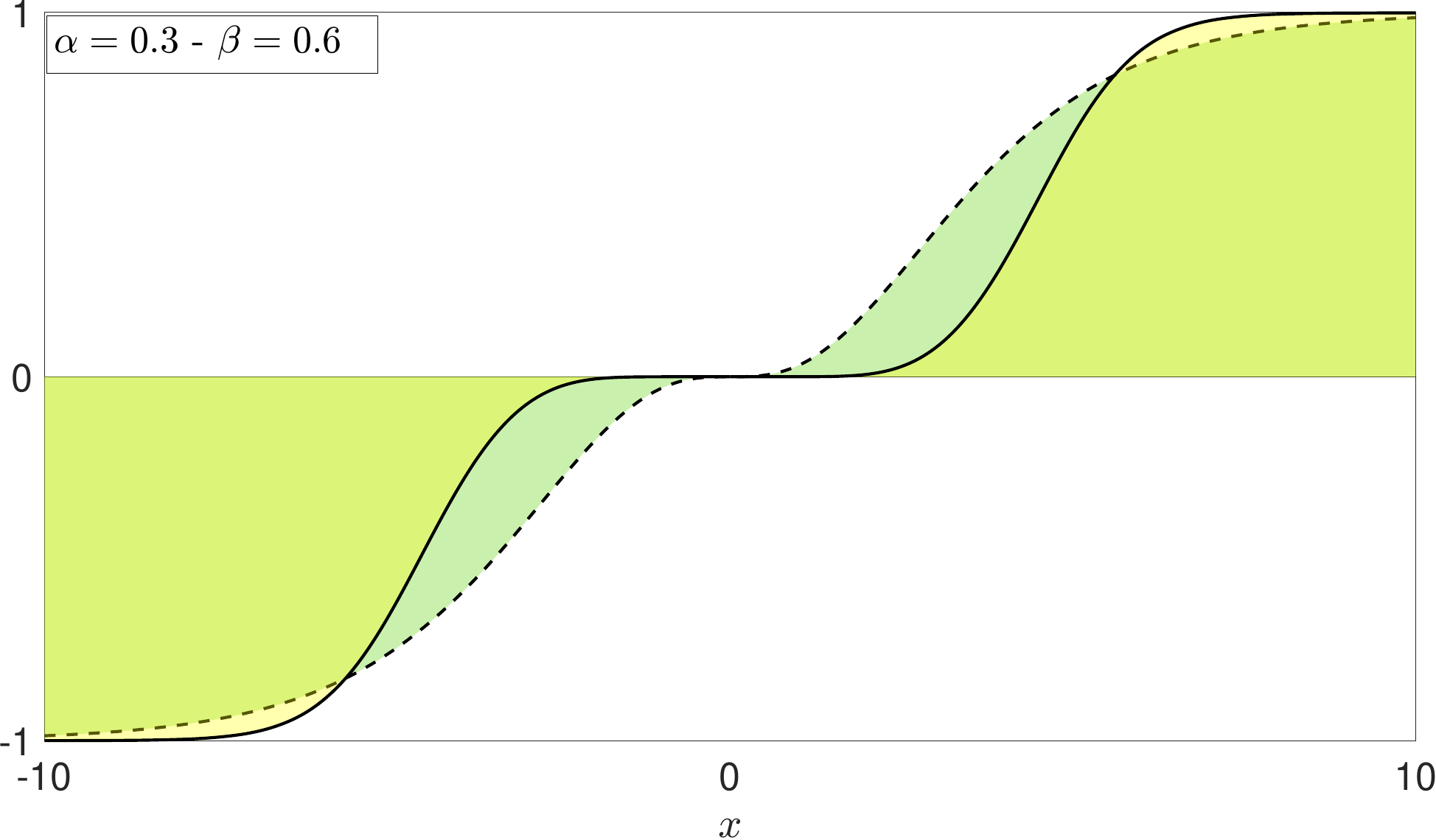} }}%
    \quad\quad\quad
    \centering{{\includegraphics[width=7.6cm]{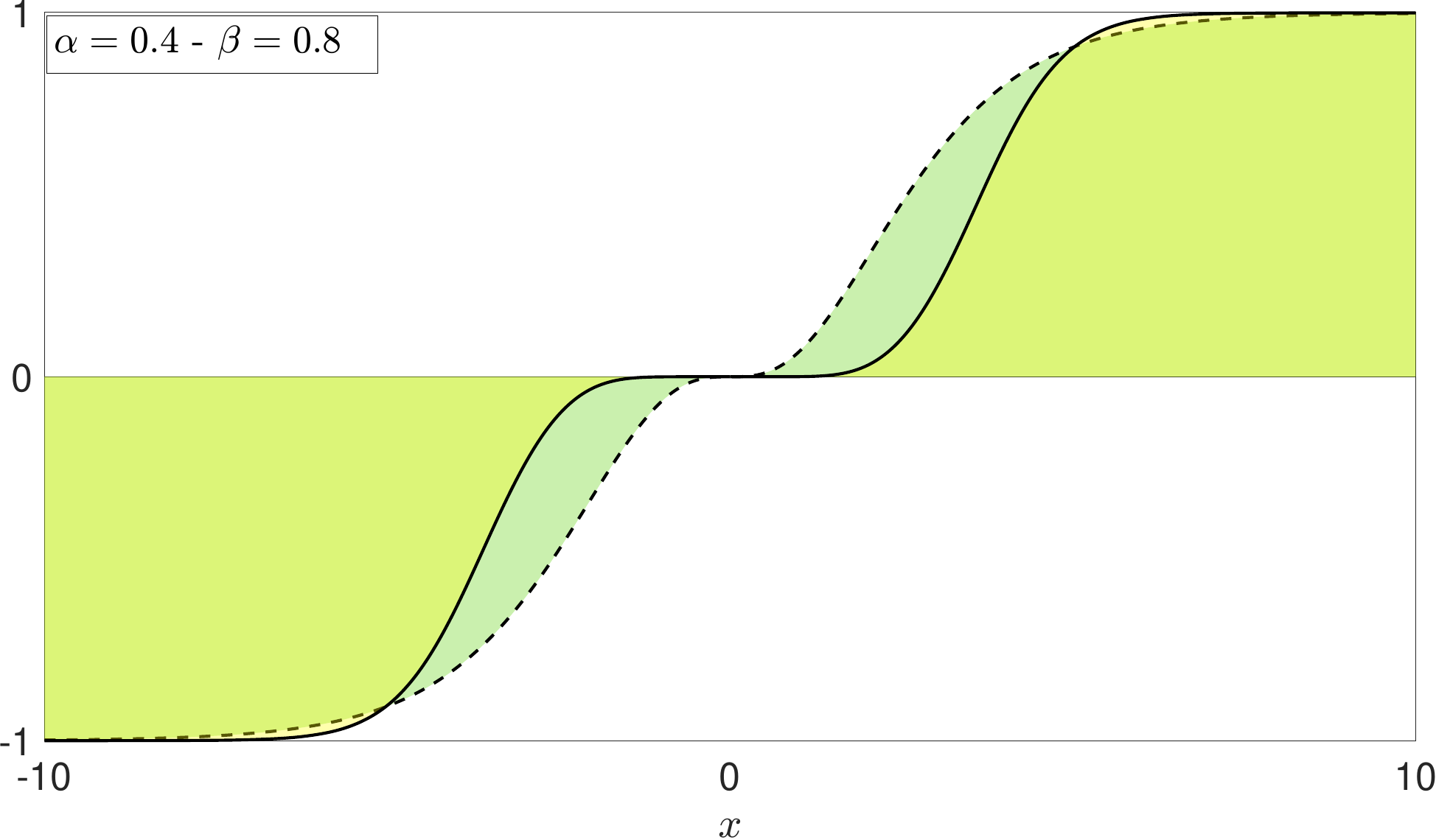} }}
\caption{The 4-4-4 model. Solutions $\phi$ (solid) and $\chi$ (dashed) for $f(\chi,\psi) = 1/(\chi\psi)^2$ and $g(\psi) = 1/\psi^2$. }\label{fig.9}
\end{figure}

This potential has minima at $\phi_{\pm} = \pm 1$, $\chi_{\pm} = \pm 1$ and $\psi_{\pm} = \pm 1$. In this model, the first-order equations \eqref{1oe} become, taking the upper signal of the equations

\begin{align}
\begin{split}
 \frac{d \psi}{dx} = \beta(1-\psi^2),
\;\;\;\;\;\;\;\;
\frac{d \chi}{dx} = \alpha\psi^2(1-\chi^2),
\;\;\;\;\;\;\;\;
\frac{d \phi}{dx} = \psi^2\chi^2(1-\phi^2).
\end{split}
\end{align}
Since we are dealing with a $\psi^4$ model, the solution was already calculated in Eq. \eqref{psi4}. Substituting this result in the first-order equation for the field $\chi$, we get the same solution obtained in \eqref{45}. Using this result in the first-order equation for the field $\phi$, we get
\begin{equation}
    \phi' = \tanh^2\PC{\alpha Y_{\beta}(x)}\tanh^2(\beta x)(1-\phi^2),
\end{equation}
which we solved numerically, giving the solutions depicted in Fig \ref{fig.9}. It has a very similar behaviour, when we compare to the first model that we analyzed in Sec. \ref{IIIA1}. Since, on the definition of $f(\chi,\psi)$, the term $1/\chi^2$ diverges inside an interval that contains the divergence of $1/\psi^2$, the addition of the previous term does not induce important modifications on the model.

\subsection{$f(\chi,\psi) = 1/\chi^2(1-\psi^2)^2$\;\;\;\;\;$g(\psi) = 1/\psi^4$}
We have that $f(\chi,\psi) = 1/\chi^2(1-\psi^2)^2$ and $g(\psi) = 1/\psi^4$. Since we are dealing with the 4-4-4 model, the auxiliary function is defined as in Eq. \eqref{W444}. Using \eqref{Potential}, the potential becomes
\begin{equation}
    V(\phi,\chi,\psi) = \frac{1}{2}\chi^2(1-\psi^2)^2\PC{1-\phi^2}^2 + \frac{1}{2}\alpha^2\psi^4\PC{1-\chi^2}^2 + \frac{1}{2}\beta^2\PC{1-\psi^2}^2.
\end{equation}

\begin{figure}%
    \centering
    \centering{{\includegraphics[width=7.6cm]{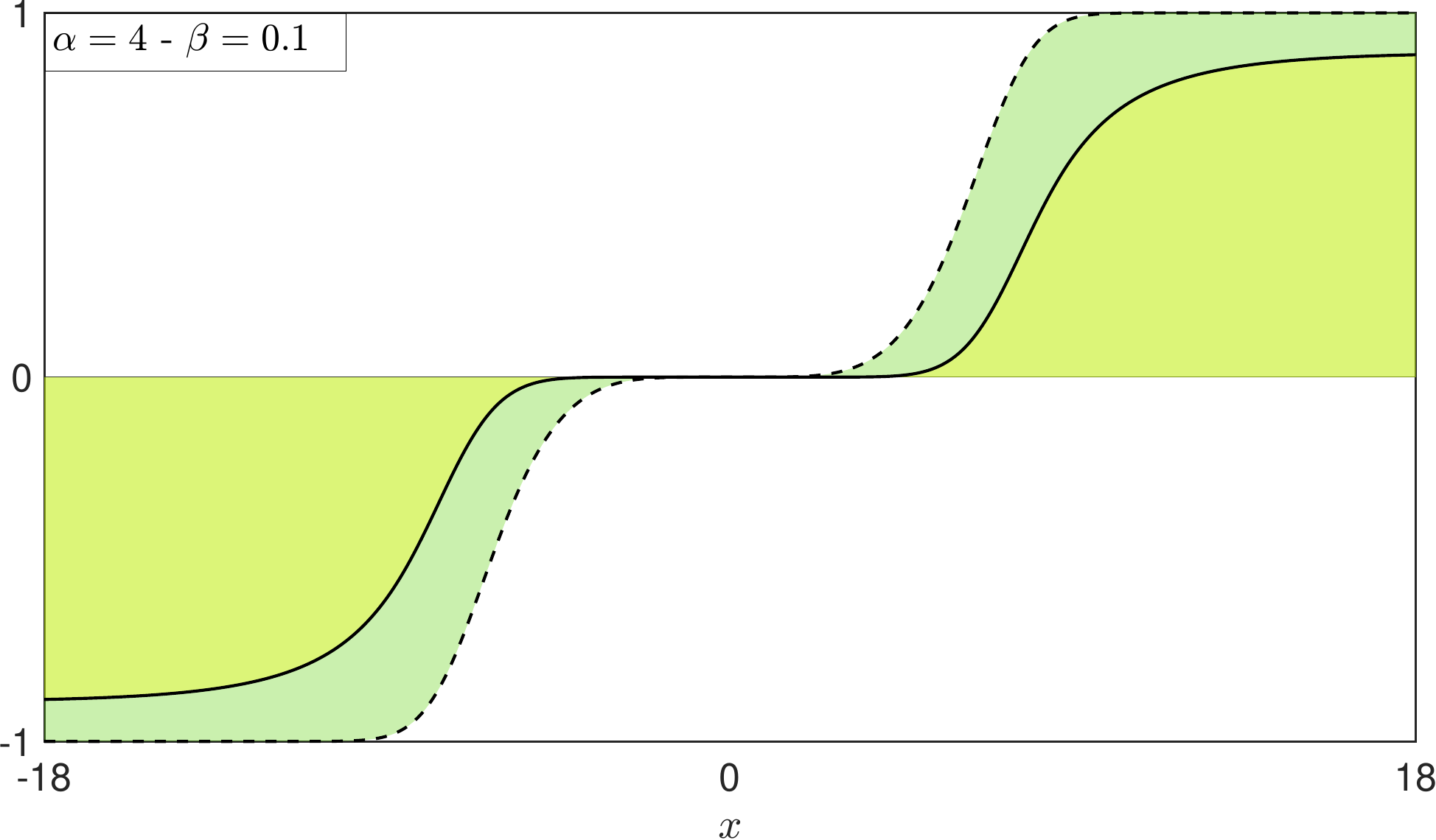} }}%
    \quad\quad\quad
    \centering{{\includegraphics[width=7.6cm]{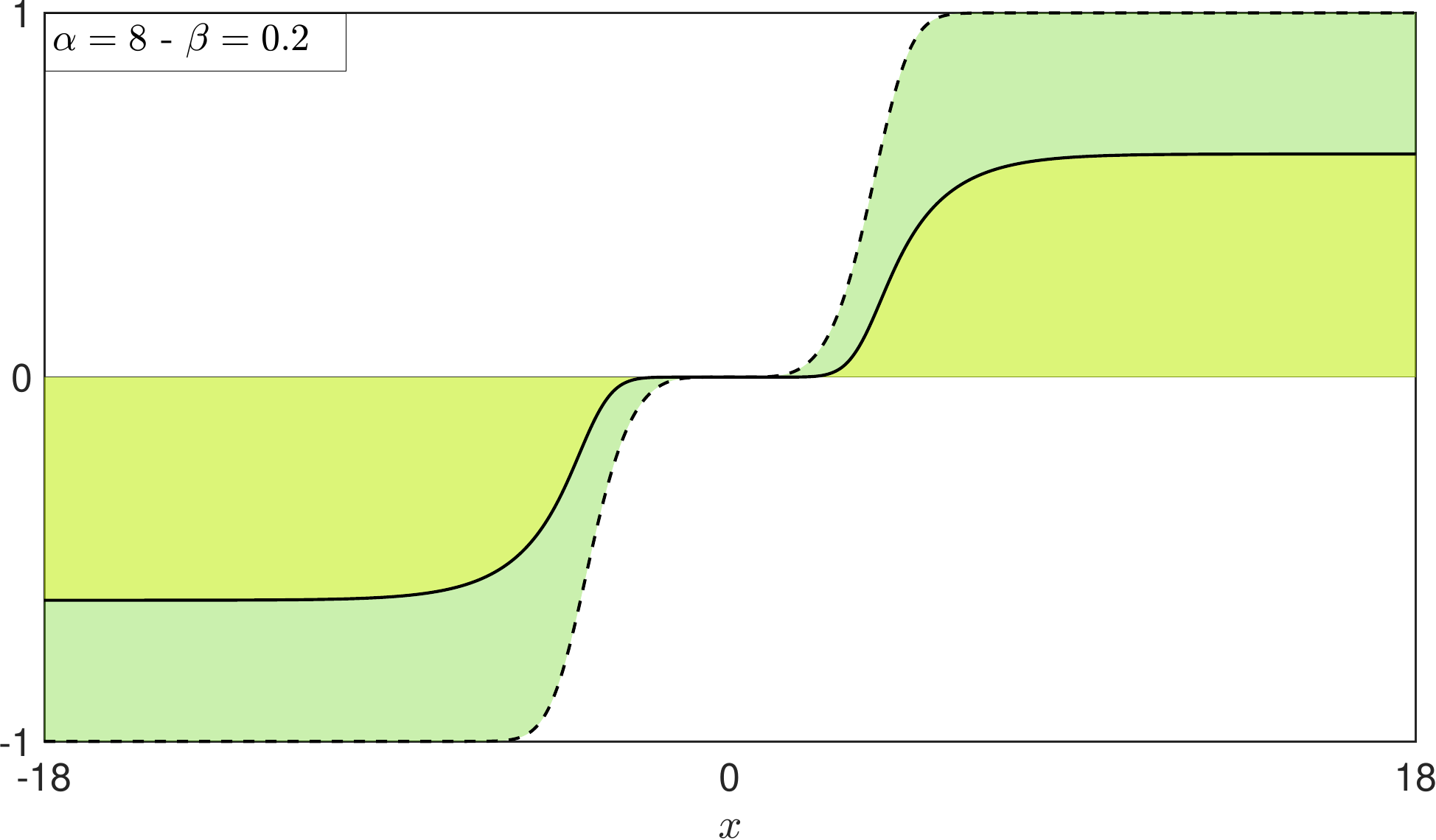} }}
\caption{The 4-4-4 model. Solutions $\phi$ (solid) and $\chi$ (dashed) for $f(\chi,\psi) = 1/\chi^2(1-\psi^2)^2$ and $g(\psi) = 1/\psi^4$. }\label{fig.10}
\end{figure}

This potential has minima at $\phi_{\pm} = \pm 1$, $\chi_{\pm} = \pm 1$ and $\psi_{\pm} = \pm 1$, and other continuum lines of minima. In this model, the first-order equations \eqref{1oe} become
\begin{align}
\begin{split}
 \frac{d \psi}{dx} = \beta(1-\psi^2),
\;\;\;\;\;\;\;\;
\frac{d \chi}{dx} = \alpha\psi^4(1-\chi^2),
\;\;\;\;\;\;\;\;
\frac{d \phi}{dx} = \chi^2(1-\psi^2)^2(1-\phi^2).
\end{split}
\end{align}
Since we are dealing with the $\psi^4$ model, the solution was already calculated in Eq. \eqref{psi4}. Substituting this result in the first-order equation for the field $\chi$, we get as solution
\begin{equation}\label{solC}
    \chi = \tanh\PR{\alpha x - \frac{\alpha}{3\beta}\tanh(\beta x)\PC{3+\tanh^2(\beta x)}}.
\end{equation}
Using this result in the first-order equation for the field $\phi$, we get
\begin{equation}
    \phi' = \sech^4(\beta x)\tanh^2\PR{\alpha x - \frac{\alpha}{3\beta}\tanh(\beta x)\PC{3+\tanh^2(\beta x)}}(1-\phi^2),
\end{equation}
which we solved numerically, giving the solutions showed in Fig \ref{fig.10}. Interestingly, the field $\phi$ has a plateau at $x = 0$, but it also has the asymptotic values changed symmetrically. This is due to the definition of $f(\chi,\psi)$, where the term $1/\chi^2$ diverges when $x=0$, which induces a internal structure. On the other hand, the term $1/(1-\psi^2)^2$ diverges only when $x\rightarrow \pm \infty$, which induces the modification on the asymptotic values of $\phi(x)$. This modification is symmetric because $\psi^2$ is also symmetric.

\subsection{$f(\chi,\psi) = 1/\chi^2(1-\psi)^2$\;\;\;\;\;$g(\psi) = 1/\psi^4$}
We have that $f(\chi,\psi) = 1/\chi^2(1-\psi)^2$ and $g(\psi) = 1/\psi^4$. Since we are dealing with the 4-4-4 model, the auxiliary function is defined as in Eq. \eqref{W444}. Using \eqref{Potential}, the potential gets to the form
\begin{equation}
    V(\phi,\chi,\psi) = \frac{1}{2}\chi^2(1-\psi)^2\PC{1-\phi^2}^2 + \frac{1}{2}\alpha^2\psi^4\PC{1-\chi^2}^2 + \frac{1}{2}\beta^2\PC{1-\psi^2}^2.
\end{equation}

\begin{figure}[h!]
    \centering
\centering{{\includegraphics[width=7.6cm]{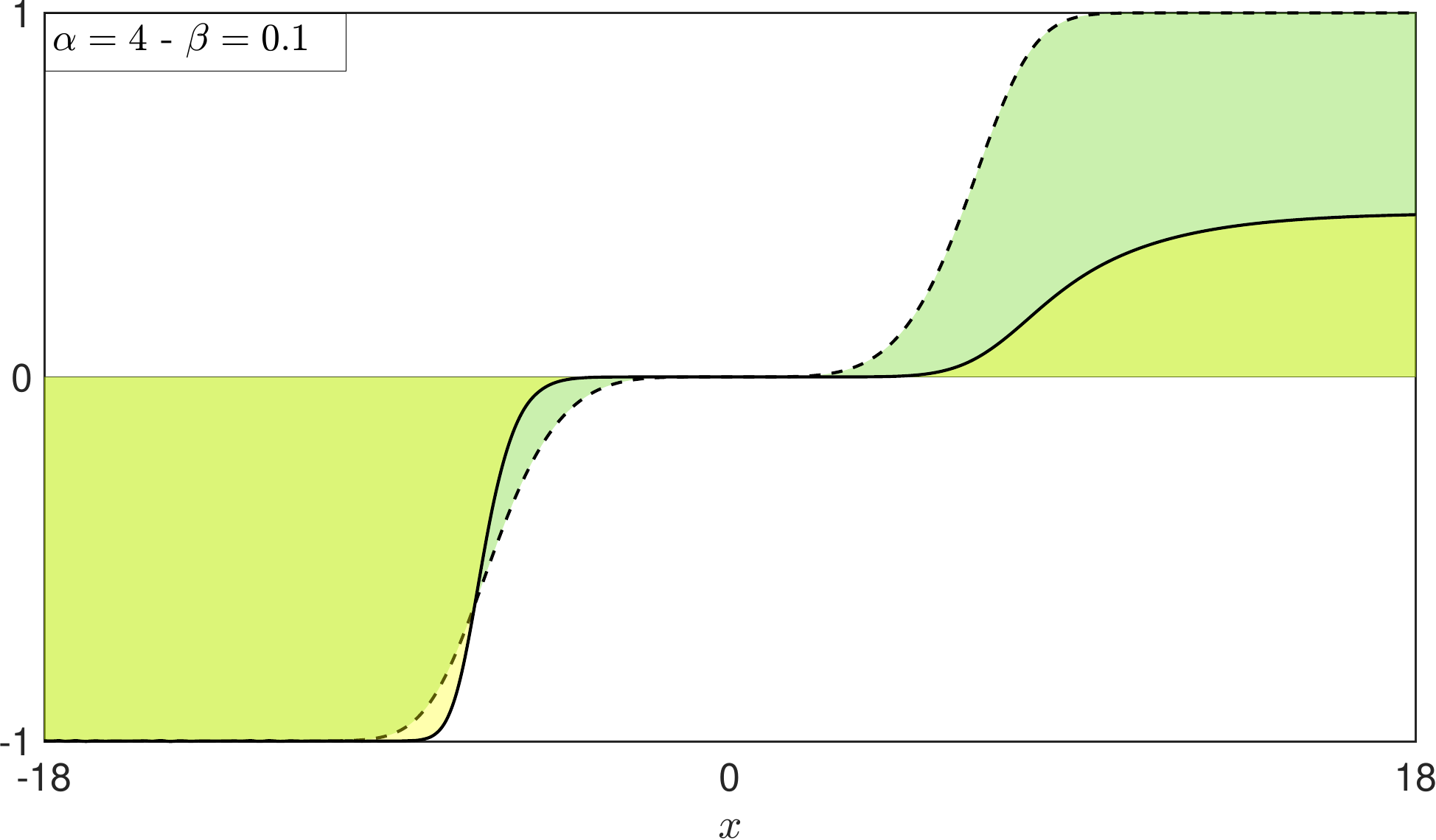} }}%
    \quad\quad\quad
\centering{{\includegraphics[width=7.6cm]{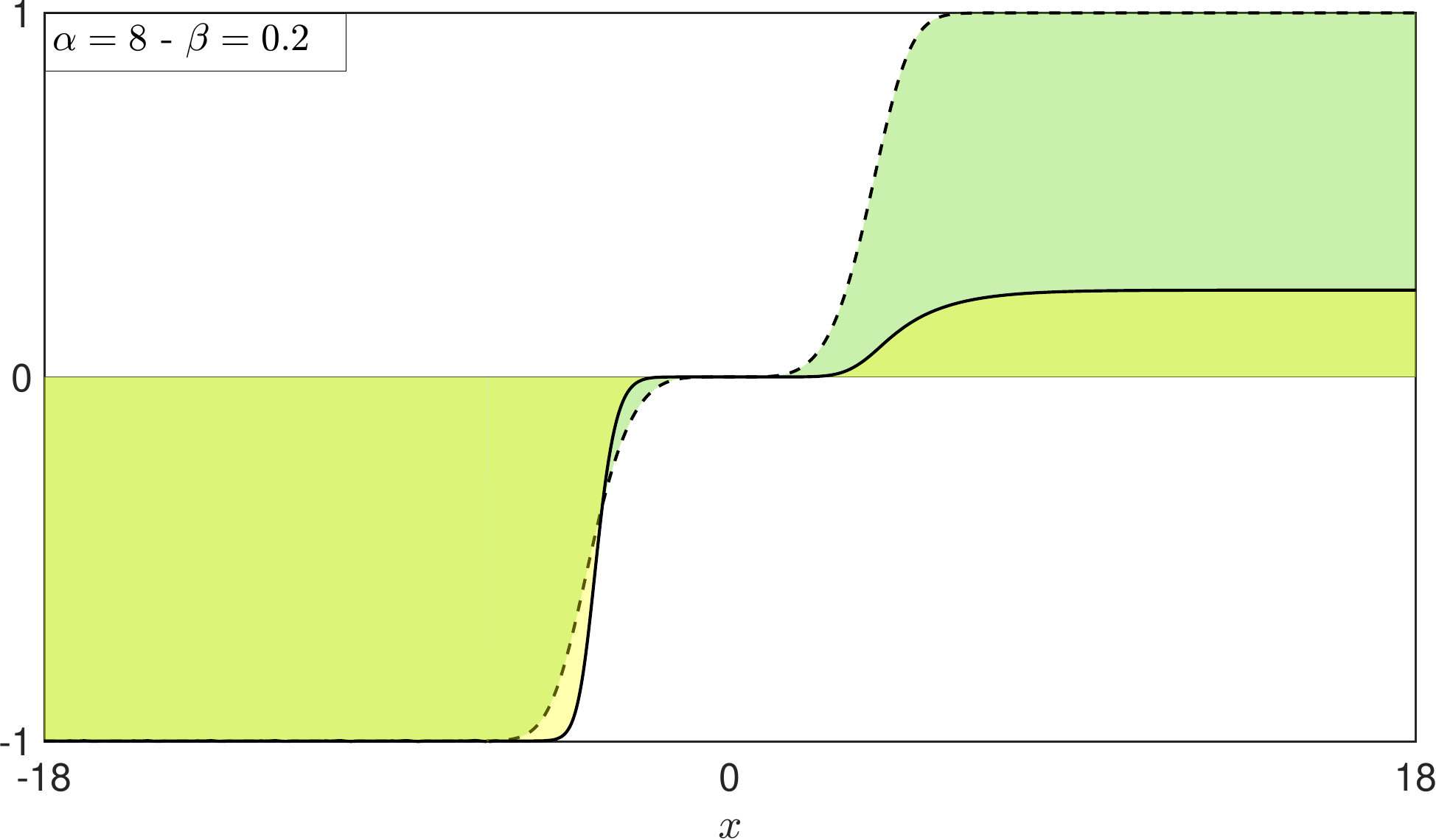} }}
\caption{The 4-4-4 model. Solutions $\phi$ (solid) and $\chi$ (dashed) for $f(\chi,\psi) = 1/\chi^2(1-\psi)^2$ and $g(\psi) = 1/\psi^4$. }\label{fig11}
\end{figure}

This potential has minima at $\phi_{\pm} = \pm 1$, $\chi_{\pm} = \pm 1$ and $\psi_{\pm} = \pm 1$, and other lines of continuum minima. In this model, the first-order equations \eqref{1oe} become
\begin{align}
\begin{split}
 \frac{d \psi}{dx} = \beta(1-\psi^2),
\;\;\;\;\;\;\;\;
\frac{d \chi}{dx} = \alpha\psi^4(1-\chi^2),
\;\;\;\;\;\;\;\;
\frac{d \phi}{dx} = \chi^2(1-\psi)^2(1-\phi^2).
\end{split}
\end{align}
Since we are dealing with the $\psi^4$ model, the solution was already calculated in Eq. \eqref{psi4}. Substituting this result in the first-order equation for the field $\chi$, we get the same solution obtained in Eq. \eqref{solC}.
Using this result in the first-order equation for the field $\phi$, we get
\begin{equation}
    \phi' = (1-\tanh(\beta x))^2\tanh^2\PR{\alpha x - \frac{\alpha}{3\beta}\tanh(\beta x)\PC{3+\tanh^2(\beta x)}}(1-\phi^2),
\end{equation}
which we solved numerically, giving the solutions displayed in Fig \ref{fig11}. Differently from the last model, the modification that $\psi$ induces on the tail of the field $\phi$ is asymmetric. This is a consequence of the term $1/(1-\psi)^2$, which only diverges for $x\rightarrow +\infty$, thus modifying the asymptotic value of $\phi(x)$ in this limit. However, if we consider $1/(1+\psi)^2$, we modify the left tail of $\phi(x)$ due to the divergence now occurring when $x\rightarrow -\infty$.

\section{Conclusions}\label{V}

In this work, we studied models described by three real scalar fields in $1+1$ spacetime dimensions. We introduced modifications in the kinetic parts of two of the three fields, to investigate how the added contributions work to change the localized structures. We first focused on the methodology, to describe the main steps to get to first-order differential equations that solve the equations of motion and give rise to stable localized configurations.  We then illustrated our findings considering two distinct families of models in the two Secs. \ref{III} and \ref{IV}, defined by specific functions $W(\phi,\chi,\psi)$, $f(\chi,\psi)$ and $g(\psi)$. The illustrations included several distinct models, and we commented on the effects added to the field $\phi$ in all specific cases considered in the two families. One interesting behavior reveled by the kinematical modification in the field $\phi$ is that, when properly included, the effects of the fields $\chi$ and $\psi$ may add together, enhancing their contribution to $\phi$. 

We think there are several perspectives of continuation of the present work, in particular, the study of other functions $W(\phi,\chi,\psi)$, $f(\chi,\psi)$ and $g(\psi)$, and the case with four or more scalar fields. We can consider adding to $f$ and $g$, contributions of the Bessel functions $J_0$ and/or $J_1$, inspired by the presence of optical solitons in Bessel optical lattices, as discussed in Refs. \cite{BL1,BL2}. Another possibility is to consider the three-field model as an extension of the two-field model investigated before in \cite{LV}. There, the authors included a term that breaks Lorentz invariance, and we believe the breaking of Lorentz invariance in high energy physics is similar to the Dzyaloshinskii-Moriya (DM) interaction in magnetic materials \cite{D,M}, which describes a magnetic exchange interaction between two neighboring magnetic spins. It is a source of weak ferromagnetic behavior of direct importance to the production of magnetic skyrmions; see, e.g., Refs.  \cite{S1,S2} and references therein for more information on skyrmions in magnetic materials. 
This line of investigation seems to be of direct interest to condensed matter, to the study of skyrmions in magnetic elements having the appropriate properties. An apparent barrier here concerns the fact that we are working in one spatial dimension, and skyrmions are two-dimensional structures. However, we can follow the lines of \cite{S3} to smoothly navigate from one to two spatial dimensions, so we believe the results of the present work can also be used to study skyrmions in magnetic materials. 

In the two-field model studied in \cite{LV}, the Lorentz violating term couples the fields $\phi$ and $\chi$ including the extra contribution $\phi k^\mu\partial_\mu\chi$, with $k^\mu$ being a constant vector. In the several three-field models discussed above, an interesting new possibility is to use the third field $\psi$ to investigate how it can geometrically constrain the fields $\phi$ and $\chi$ in the presence of the above LV term. Another issue related to this concerns the case of spectral walls uncovered sometime ago in \cite{SW1}. The point is that in the more recent work \cite{SW2}, the authors considered a two-field model with the inclusion of a distinct Lorentz breaking term, to investigate spectral wall properties. This study has shown that the presence of Lorentz breaking is an interesting ingredient to unveil the spectral wall phenomenon, in this sense suggesting that some spectral wall effects may also appear for skyrmions in magnetic materials guided by the DM interaction.

Another possibility of interest in application in magnetic materials can be related to the breaking of chirality in nonrelativistic nonequilibrium systems similar to the X-Y model considered before in Ref. \cite{XY}, where a transition between Ising and Bloch walls was found to be possible, under the introduction of a new term that controls the presence of weak anisotropy in the system. This mechanism is distinct from the LV contribution and is worth being investigated. In particular, we can add another real scalar field to the model used in \cite{XY} to extend the system to the three real scalar fields framework studied in the present work.

The three field models can be considered in a way similar to some recent investigations, to see how the scattering processes studied in \cite{Campos} work for three fields, and if it may involve topological charge exchange and kink–antikink bound state formation \cite{PhysD1}, the interaction between the shape modes \cite{PhysD3} and also, the study of kink in non-linear Sigma models with the target space being the torus $S^1 \times S^1$ \cite{physD2}. They can also be used to describe braneworld scenarios \cite{BW1,BW2,BW21,BW3,BW4}, with the kinematical modifications changing the internal structure of the brane, in a way similar to the investigation described more recently in \cite{34,BW5}. We can also consider periodic potentials like in the sine-Gordon model \cite{B2}, in the nonintegrable sine-Gordon model with the presence of internal modes \cite{SG1}, generalization of the sine-Gordon model to the case of two fields \cite{SG2} and the modified model supporting multi-kink configurations \cite{SG3}. These and other related issues are presently under consideration, and we hope to report on them in the near future. 

\acknowledgments{
This work was partially financed by Coordenação de Aperfeiçoamento de Pessoal de Nível Superior (CAPES), Grant 88887.899555/2023-00 (GSS), by Conselho Nacional de Desenvolvimento Científico e Tecnológico (CNPq), Grant 303469/2019-6 (DB), and by Paraiba State Research Foundation, Grant 0015/2019 (DB).
}


\end{document}